\documentclass{aa} 
\usepackage{txfonts,textcomp}

\usepackage{graphicx}
\usepackage{placeins}
\usepackage[inline]{enumitem}
\usepackage{tabulary}
\usepackage{booktabs}
\usepackage{orcidlink}
\newcommand\HII{\ion{H}{II}\xspace} 
\newcommand\NII{[\ion{N}{II}]\xspace} 
\newcommand\OIII{[\ion{O}{III}]\xspace} 

\newcommand\CII{\ion{C}{II}\xspace} 
\newcommand\CIIforb{[\ion{C}{II}]\xspace} 
\newcommand\OI{[\ion{O}{I}]\xspace} 
\newcommand\SIII{[\ion{S}{III}]\xspace} 
\setcitestyle{citesep={,}}

\defcitealias{2023MNRAS.526.3871K}{Paper I}
\defcitealias{2021MNRAS.506.5703K}{K21}

\begin{document}
\titlerunning{Mock observations with \texttt{TODDLERS}}
\title{\texttt{TODDLERS}: A new UV-millimeter emission library for star-forming regions} 
\subtitle{II. Star-formation rate indicators using Auriga zoom simulations}
\author{
Anand Utsav Kapoor\inst{1}\orcidlink{0000-0002-5187-1725}\thanks{E-mail: anandutsavkapoor@gmail.com} \and
Maarten Baes\inst{1}\orcidlink{0000-0002-3930-2757} \and
Arjen van der Wel\inst{1}\fnmsep\orcidlink{0000-0002-5027-0135} \and
Andrea Gebek\inst{1}\orcidlink{0000-0002-0206-8231} \and
Peter Camps\inst{1}\fnmsep\orcidlink{0000-0002-4479-4119} \and
Aaron Smith\inst{2}\orcidlink{0000-0002-2838-9033} \and
Médéric Boquien\inst{4} \and
Nick Andreadis\inst{1} \and
Sebastien Vicens\inst{3}
}

\institute{
Sterrenkundig Observatorium, Universiteit Gent, Krijgslaan 281-S9, B-9000 Gent, Belgium
\and
The University of Texas at Dallas, 800 W Campbell Rd, Richardson, TX 75080, USA
\and
Faculté des sciences et de génie, Université Laval, Pavillon Alexandre-Vachon, Québec, G1V 0A6, Canada
\and
Université Côte d'Azur, Observatoire de la Côte
d'Azur, CNRS, Laboratoire Lagrange, F-06000 Nice, France
}

\date{Accepted XXX. Received YYY; in original form ZZZ}




\abstract
{The current generation galaxy formation simulations often approximate star formation, making it necessary to use models of star-forming regions to produce observables from such simulations. In the first paper of this series, we introduced \texttt{TODDLERS}, a physically motivated, time-resolved model for UV--millimeter (mm) emission from star-forming regions, implemented within the radiative transfer code \texttt{SKIRT}. In this work, we use the \texttt{SKIRT-TODDLERS} pipeline to produce synthetic observations.}
{We aim to demonstrate the potential of \texttt{TODDLERS} model through observables and quantities pertaining to star-formation. An additional goal is to compare the results obtained using \texttt{TODDLERS} with the existing star-forming regions model in \texttt{SKIRT}}
{We calculated broadband and line emission maps for the 30 Milky Way-like galaxies of the Auriga zoom simulation suite at a redshift of zero. Analyzing far-ultraviolet (FUV) and infrared (IR) broadband data, we calculated kiloparsec (kpc)-resolved IR correction factors, \(k_{IR}\), which allowed us to quantify the ratio of FUV luminosity absorbed by dust to reprocessed IR luminosity. Furthermore, we used the IR maps to calculate the kpc-scale mid-infrared (MIR) colors ($8\,\mu \rm{m} / 24\,\mu \rm{m}$) and far-infrared (FIR) colors ($70\,\mu \rm{m} / 500\,\mu \rm{m}$) of the Auriga galaxies.
We used H$\alpha$ and H$\beta$ line maps to study the Balmer decrement and dust correction. We verified the fidelity of our model's FIR fine structure lines as star formation rate (SFR) indicators.}
{The integrated UV-mm spectral energy distributions (SEDs) exhibit higher FUV and and near-ultraviolet (NUV) attenuation and lower $24\,\mu \rm{m}$ emission compared to the existing star-forming regions model in \texttt{SKIRT}, alleviating tensions with observations reported in earlier studies. The light-weighted mean \(k_{IR}\) increases with aperture and inclination, while its correlation with kpc-resolved specific star-formation rate (sSFR) is weaker than literature values from resolved SED fitting, potentially due to inaccuracies in local energy balance representation. The kpc-scale MIR-FIR colors show an excellent agreement with local observational data, with anti-correlation degree varying by galaxy morphology. We find that the Balmer decrement effectively corrects for dust, with the attenuation law varying with dust amount. The H$\alpha$ emission attenuation levels in our models are comparable to those observed in the high-density regions of state-of-the-art radiation hydrodynamical simulations. The FIR fine-structure line emission-based luminosity-SFR relations are consistent with global observational relations, with the \CIIforb\ line displaying the best agreement.}
{}

\keywords{Galaxies: ISM --
    Radiative transfer -- 
    Methods: numerical -- 
    HII regions -- 
    ISM: lines and bands -- 
    (ISM:) dust, extinction
}


\maketitle

\section{Introduction}

Star formation is a fundamental phenomenon in astrophysics, crucial for shaping the evolution of galaxies \citep{1998ARA&A..36..189K}. This process transforms dust and gas within molecular clouds into stars, injecting chemical elements and energy into the interstellar medium \citep[ISM;][]{2007ARA&A..45..565M, 2014PhR...539...49K, 2020SSRv..216...50C}. Young, massive stars at star-forming sites play a crucial role in shaping a galaxy's spectral energy distribution (SED), producing some of the most prominent observational signatures across wavelengths \citep{2006ApJ...649..759C, 2007ApJ...670..428C, 2019PASJ...71....6H,2023MNRAS.522.6236T}. Emission from recent star formation dominates the UV output of galaxies \citep{2007ApJS..173..267S, 2007ApJS..173..185G, 2011ApJ...741..124H, 2011MNRAS.413.2570R}, while reprocessed emission by dust creates prominent infrared (IR) features \citep{2000ApJ...533..682C, 2009ApJ...692..556R, 2012ARA&A..50..531K}. Furthermore, massive stars emit radiation that ionizes the surrounding gas, leading to the production of recombination emission lines such as H$\alpha$ and H$\beta$ \citep{1999ApJS..123....3L,2006agna.book.....O}. These lines (in)directly trace the ionizing radiation output of young, massive stars. The energetic free electrons produced through ionization heat the gas, and the subsequent cooling through metal emission also leaves important observational signatures \citep{1985ApJS...57..349R, 1998PASP..110..761F}. These emission features serve as crucial tools for uncovering the physical conditions and chemical makeup of the ISM, providing valuable insights into galaxy formation and evolution \citep{2002ApJS..142...35K, 2019ARA&A..57..511K, 2022MNRAS.513.5134N, 2023ARA&A..61..473C}.

On the theoretical front, hydrodynamical numerical simulations have become indispensable in the study of galaxy formation and evolution \citep{2015ARA&A..53...51S, 2020NatRP...2...42V}. The scale of the model significantly influences its focus and complexity, necessitating the use of sub-grid models to approximate unresolved phenomena such as star formation, stellar feedback, and the interaction of energy and matter on scales smaller than can be directly simulated.
Currently, some state-of-the-art isolated galaxy simulations are able to achieve a parsec-scale (pc-scale) resolution, allowing them to resolve cold molecular gas in regions where stars are born. Notable works in this direction include \citet{2018MNRAS.478..302S, 2019MNRAS.489.4233M,10.1093/mnras/staa3249,2023MNRAS.522.3831F, 2022arXiv221104626K, 2022MNRAS.517.1557R, 2024ApJ...960..100S}, and \citet{2024arXiv240508869D}. Although the detailed processes of star formation are realized through sub-grid models, these high-resolution simulations represent the most realistic current models of the influence of stellar feedback on the surrounding gas in galaxy simulations. On the other hand, simulations that encompass a larger number of galaxies enable the study of galaxy formation and evolution in a cosmological context. These efforts generally come in two guises. Some directly simulate a large co-moving volume, such as Illustris \citep{2014MNRAS.444.1518V}, EAGLE \citep{2017MNRAS.464.4204C}, Simba \citep{2019MNRAS.486.2827D}, IllustrisTNG \citep{2019MNRAS.490.3196P}, and NewHorizon \citep{2021A&A...651A.109D}. Others zoom in on interesting regions or halos with specific characteristics, using a higher number of resolution elements, such as NIHAO \citep{2015MNRAS.454...83W}, Auriga \citep{2017MNRAS.467..179G}, ARTEMIS \citep{2020MNRAS.498.1765F}, THE THREE HUNDRED project \citep{2022MNRAS.514..977C}, and HELLO project \citep{2024MNRAS.533.1463W}.
In both cases, due to limitations in the physical modeling, resolution constraints, or the high computational cost from very small time steps in dense gas regions, approximate treatments of the ISM are necessary. These approximations often result in an ISM that is highly unresolved and overly smooth. Common techniques include implementing an effective equation of state and/or artificial temperature as well as pressure floors to prevent unresolved numerical fragmentation and maintain computational stability \citep{2003MNRAS.339..289S, 2015MNRAS.446..521S, 2019MNRAS.488.4400I}.
While these approaches enable the simulation of larger volumes or higher resolution in specific regions, they significantly impact the modeled ISM structure, limiting the ability to resolve its cold, dense phases. Furthermore, the use of such methods fundamentally changes the behavior of mechanical, chemical, radiative, and thermodynamical feedback processes. 
As a result, simulated galaxies frequently display smoother structures on scales of several hundred parsecs and tend to be thicker and kinematically hotter than observed \citep{2019MNRAS.489.4233M, 2018MNRAS.473.1019B}. Thus, these simulations are unable to capture detailed small-scale information about star-forming environments, including the intricate physical conditions such as gas density, temperature, and chemical composition. These gas conditions are critical for determining the emission from star-forming regions as they influence the ionization states and excitation mechanisms \citep{2006agna.book.....O, 2011piim.book.....D, ryden_pogge_2021}. 

While efforts to simulate the multi-phase ISM continue for the next generation of cosmological simulations \citep[see, e.g., ][]{2020MNRAS.497.4857P, 2023MNRAS.523.3709C}, the production of synthetic observations using current-generation large-volume simulations requires sub-grid models to incorporate small-scale gas emission, such as those from star-forming regions. This helps ensure a proper accounting of the complex interplay of physical processes at smaller scales. Such an approach bridges the gap between large-scale cosmological simulations and the small-scale information required for predicting emissions from star-forming regions. Several examples in the literature use this method to predict individual line luminosities \citep{2017ApJ...846..105O, 2020MNRAS.496..339P, 2021ApJ...922...88O, 2023MNRAS.526.3610H, 2022ApJ...926...80G}. However, there is a dearth of models that predict both continuum and line-emission from UV to millimeter wavelengths as this requires combining multiple specialty radiative transfer techniques. Our work is dedicated to such multi-wavelength modeling in the context of high-resolution hydrodynamical simulations. By adopting this approach, we aim to enhance the predictive power of current-generation simulations, striving for greater realism while maintaining similar computational efforts.

In \citet[][hereafter \citetalias{2023MNRAS.526.3871K}]{2023MNRAS.526.3871K}, we introduced a physically motivated, time-resolved representation of UV to millimeter (UV--mm) emission from star-forming regions, including both emission lines and continua. The model was constructed using a 1D evolution model for a spherical, homogeneous gas cloud exposed to stellar feedback, coupled with the photoionization code \texttt{Cloudy}. The inclusion of the \texttt{TODDLERS} library in \texttt{SKIRT}, a state-of-the-art three-dimensional (3D) multi-physics Monte Carlo radiative transfer code, facilitates forward modeling of cosmological hydrodynamical simulations. This integration provides meaningful insights from the complex observations made by integral field unit (IFU) instruments. This is particularly relevant and timely given the increasing number of advanced IFU instruments  such as  Multi Unit Spectroscopic Explorer \citep[MUSE;][]{2010SPIE.7735E..08B} on the Very Large Telescope (VLT),  Atacama Large Millimeter/submillimeter Array \citep[ALMA;][]{5136193},  James Webb Space Telescope \citep[JWST;][]{2006SSRv..123..485G}, and  Spectromètre Imageur à Transformée de Fourier pour l'Etude en Long et en Large de raies d'Emission \citep[SITELLE;][]{2019MNRAS.485.3930D} on the  Canada-France-Hawaii Telescope (CFHT) are actively gathering data, allowing for spectrally and spatially resolved observations, enabling comprehensive mapping of galaxy properties. While \citetalias{2023MNRAS.526.3871K} laid the foundations of this investigation by detailing the methodology, key diagnostics, and comparisons with state-of-the-art models (such as the one presented in \citet{2008ApJS..176..438G}, referred to as \texttt{HiiM3} for the version available in \texttt{SKIRT}), this work focuses on applying the \texttt{TODDLERS} library to generate synthetic observables using the Auriga suite of simulations \citep{2017MNRAS.467..179G}. Through an exploration of broadband images, optical, and FIR spectra, we aim to demonstrate the capabilities and potential of the \texttt{TODDLERS} model and provide a detailed comparison of the synthetic data with observations to validate our approach.

In Sect.~\ref{sect:background}, we provide background information on the Auriga simulations, the DustPedia project, the \texttt{SKIRT} radiative transfer code, and the \texttt{TODDLERS} model, which comprise the tools and datasets used in our study. In Sect.~\ref{sect:methodology}, we describe the process of exporting Auriga snapshots to \texttt{SKIRT} for performing the RT simulations and the calibration of the free parameters for accurate physical modeling.
Section~\ref{sect:data_products} describes the data products generated in this work.
In Sect.~\ref{sec:UV-mm SED and broadband data analysis}, we compare the Auriga SEDs obtained using \texttt{TODDLERS} in the post-processing pipeline with those obtained using \texttt{HiiM3}. We also analyze the broadband images generated in this work, focusing on hybrid UV-IR star-formation rate (SFR) indicators and mid-infrared-far-infrared (MIR-FIR) colors to evaluate their utility and correspondence to observed galaxy properties. In Sect.~\ref{sect:emission_line_based_SF_maps}, we discuss and present the analysis of the Balmer and FIR fine-structure line emission of Auriga galaxies using \texttt{TODDLERS}.
Finally, in Sect.~\ref{sec:summary}, we summarize our findings, present our conclusions, and suggest directions for future research and applications.

\section{Background}
\label{sect:background}

\subsection{Auriga simulations}
\label{subsect:auriga}

Auriga \citep{2017MNRAS.467..179G, 2024MNRAS.532.1814G} is a set of 30 zoom simulations aimed at the modeling of Milky Way-type galaxies in a full cosmological context, carried out using the moving-mesh magnetohydrodynamics (MHD) code \texttt{AREPO} \citep{2010MNRAS.401..791S}. These simulations follow a $\Lambda$ cold dark matter cosmology consistent with the \citet{2014} data release. The host dark matter halos of the zoomed galaxy simulations were drawn from a dark-matter-only  simulation of a comoving side length of 100 cMpc, with selection criteria on the mass and the isolation of the host halos. The virial mass ranges between 1 and 2 times $10^{12}~{\text{M}}_{\odot}$, consistent with recent determinations of the Milky Way mass \cite[see][and references therein]{2015MNRAS.453..377W}. 

The Auriga physics model uses primordial and metal-line cooling with self-shielding corrections. A spatially uniform UV background field \citep{2009ApJ...703.1416F} is employed. 
The ISM is modeled with a two-phase equation of state from \citet{2003MNRAS.339..289S}. Star formation proceeds stochastically in gas with densities higher than a threshold density ($n_{\text{thr}} = 0.13~{\text{cm}}^{-3}$). The star formation probability in the candidate gas cells scales exponentially with time, with a characteristic time scale of $t_{\text{SF}}=2.2~\text{Gyr}$.
The single stellar population (SSP) of each star particle is represented by the \citet{2003PASP..115..763C} initial mass function (IMF).
Mass and metal returns from Type Ia supernovae (SNIa), asymptotic giant branch (AGB), and Type II supernovae (SNII) stars are calculated at each time step and are distributed among nearby gas cells with a top-hat kernel. 
The number of SNII events equals the number of
of stars in an SSP that lie in the mass range 8-100~${\text{M}}_\odot$.
The model includes gas accretion by black holes, with active galactic nucleus (AGN) feedback in radio and quasar modes, both of which are always active and are thermal in nature.
Magnetic fields are treated with ideal MHD following \cite{2013MNRAS.432..176P}.

\subsection{DustPedia project}
\label{subsect:DustPedia}

The DustPedia galaxy sample \citep{2017PASP..129d4102D} contains 875 nearby galaxies with matched aperture photometry in more than 40 bands from UV to millimeter wavelengths \citep{2018A&A...609A..37C}. 
The DustPedia galaxy sample covers a broad dynamic range of various physical properties, including stellar mass, SFR, and morphological stage, making it representative for galaxies in the local Universe. The observed and inferred physical properties of the DustPedia galaxies have studied in detail using various techniques \citep[e.g.,][]{2017A&A...605A..18C, 2020A&A...633A.100C, 2022A&A...668A.130C, 2018A&A...620A.112B, 2019A&A...623A...5D, 2019A&A...624A..80N, 2020A&A...637A..25N}. The sample is therefore ideal for a comparison between observed and simulated galaxies at $z\sim0$, as has been used in this way by various authors \citep{2020MNRAS.494.2823T, 2021MNRAS.506.5703K, 2022MNRAS.512.2728C}.
Here we follow the same approach as in \citet[hereafter referred to as \citetalias{{2021MNRAS.506.5703K}}]{2021MNRAS.506.5703K}, using the DustPedia data set to calibrate the RT post-processing recipe for the simulated Auriga galaxies (as described in Sect.~\ref{subsect:calibration}).

\subsection{\texttt{SKIRT} radiative transfer code}
\label{subsect:SKIRT}

The 3D Monte Carlo radiative transfer code \texttt{SKIRT} \citep{2011ApJS..196...22B, 2015A&C.....9...20C, 2020A&C....3100381C}
is capable of importing the output from various kinds of hydrodynamical simulations. It contains a hybrid parallelization strategy \citep{2017A&C....20...16V}, a library of flexible input models \citep{2015A&C....12...33B}, and a suite of advanced spatial grids for discretizing the medium \citep{2013A&A...560A..35C, 2014A&A...561A..77S, 2024A&A...689A..13L}. It simulates the absorption, scattering, and thermal emission by dust grains, including stochastic heating \citep{2015A&A...580A..87C} and support for kinematics \citep{2020A&C....3100381C, 2023MNRAS.524..907B} and polarisation \citep{2017A&A...601A..92P, 2021A&A...653A..34V}. Apart from dust radiative transfer, the code is designed to perform Lyman-$\alpha$ radiative transfer \citep{2021ApJ...916...39C}, X-ray radiative transfer \citep{2023A&A...674A.123V}, 21~centimeter (cm) hydrogen radiative transfer \citep{2023MNRAS.521.5645G}, and non-local thermodynamic equilibrium (non-LTE) line radiative transfer \citep{2023A&A...678A.175M} without any constraints on geometrical complexity.

The code has been extensively used to generate line-emission maps, synthetic UV to submm broadband images, spectral energy distributions, and polarisation maps for idealized galaxies \citep{2010MNRAS.403.2053G, 2017A&A...601A..92P, 2021MNRAS.507.2755L}, as well as for high-resolution 3D galaxy models \citep{2014A&A...571A..69D, 2019MNRAS.487.2753W, 2020A&A...637A..25N} and for galaxies extracted from cosmological simulations \citep{2016MNRAS.462.1057C, 2020MNRAS.492.5167V, 2021MNRAS.506.5703K, 2021A&A...653A..34V, 2022MNRAS.516.3728T, 2022MNRAS.512.2728C, 2023MNRAS.524..907B, 2024A&A...683A.181B}.

\subsection{\texttt{TODDLERS} star-forming regions emission library}\label{subsect:TODDLERS}

{T}ime evolution of {O}bservables including {D}ust
{D}iagnostics and {L}ine {E}mission from {R}egions containing young {S}tars, \texttt{TODDLERS}, was described in \citetalias{2023MNRAS.526.3871K} as a UV--mm emission library  generated by assuming the spherical evolution of a homogeneous gas cloud around a young stellar cluster that accounts for stellar feedback processes. This includes stellar winds, supernovae, and radiation pressure, as well as the gravitational forces on the gas. 
The tug-of-war between stellar feedback and gravitational force in these models generally results in one of two scenarios:
\begin{enumerate}
    \item Dissolution of the gas cloud when stellar feedback is able to unbind the system. 
    \item Gas recollapse under gravitational forces when stellar feedback isn't strong enough. In such cases, a subsequent burst of star formation using the leftover gas is assumed to occur with the same star-formation efficiency. This process can repeat itself if the combined stellar feedback from all generations of clusters is not enough to unbind the cloud.
\end{enumerate}

We note that these models are calculated assuming a burst star-formation with a Kroupa IMF in the mass range 0.1-100 $\text{M}_{\odot}$ and employ the high mass-loss Geneva tracks in the \texttt{STARBURST99} population synthesis code. 
These stellar models do not account for binary evolution or single star rotation, which lead to more homogeneous stellar evolution and affect their radiative output and spectral hardness \citep{2012RvMP...84...25M, 2018MNRAS.477..904X}. Additionally, mass loss from massive stars—an essential input to our calculations, influencing mechanical luminosity and ram pressure—remains an area of active research \citep[see, for example,][]{2021A&A...646A.180K, 2023A&A...676A.109B, 2023Galax..11..105J}. As a result, the uncertainties in the stellar models are propagated into our models.
An upcoming iteration of \texttt{TODDLERS} (Kapoor et al., in prep) will consider binary evolution using BPASS \citep{2017PASA...34...58E} and relax the burst star-formation assumption.

This semi-analytical evolution model is coupled with the photoionization code \texttt{Cloudy} \citep{2017RMxAA..53..385F} to calculate time-dependent UV--mm SEDs for star-forming clouds of varying metallicity ($Z$, same value for the gas and the stellar system/s), star-formation efficiency ($\epsilon_{\rm{SF}}$), birth-cloud density ($n_{\rm{cl}}$), and  birth-cloud mass ($M_{\rm{cl}}$). The calculated SEDs include the stellar, nebular, and dust continuum emission along with a wide range of emission lines originating from \HII, photo-dissociation, and molecular gas regimes tabulated at an intrinsic resolution of $R= \lambda / \Delta \lambda =5\times10^4$. 

\section{Methodology}
\label{sect:methodology}

In this paper, we use the new emission library \texttt{TODDLERS} to generate synthetic observations.
This is achieved by post-processing thirty $z=0$ Milky Way-like galaxies of the Auriga simulation suite. The post-processing procedure takes an approach similar to the one discussed in \citetalias{2021MNRAS.506.5703K}, changing only the sub-grid star-forming regions model with respect to \citetalias{2021MNRAS.506.5703K}, namely, by incorporating \texttt{TODDLERS} in the post-processing pipeline instead of \texttt{HiiM3}. Here, we briefly discuss the key aspects of our post-processing pipeline.

The Auriga-\texttt{SKIRT} post-processing pipeline requires the specification of two key components: the diffuse dust distribution and the radiation sources in the simulated galaxies. Regarding the diffuse dust, it is assumed that a constant fraction of the metals, denoted as $f_{\text{dust}}$, within the dust-containing interstellar medium (DISM) is present in the form of dust grains. The source geometry is derived from the stellar particle data obtained from the simulation snapshot. Suitable SEDs are assigned to these sources. During the assignment of SEDs, a distinction is made between young and old stars based on their recorded ages within the simulation.
The following presents a summary of each of these components.

\subsection{Diffuse dust spatial distribution and grain model}

For each gas cell, we set the dust density, $\rho_{\text{dust}}$, as follows:
\begin{equation}
\rho_{\text {dust }}=\left\{\begin{array}{ll}
f_{\text {dust }} Z \,\rho_{\text {gas }} & \text { if DISM, } \\
0 & \text { otherwise, }
\end{array}\right.
\label{dust_allocation_crit_pap3.eqn}
\end{equation}
where $Z$ and $\rho_{\text{gas}}$ represent the metallicity and gas density given by the gas cell's properties in the Auriga snapshot. We define the DISM by distinguishing rotationally supported interstellar gas, settled in the disk, from the hot circumgalactic gas following \citet{2012MNRAS.427.2224T}:
  \begin{equation}
   {\text{DISM}} \iff \log \left(\frac{T}{\text{K}}\right)<6+0.25 \log \left(\frac{\rho_{\text{gas}}}{10^{10}~h^2~{\text{M}}_{\odot}~{\text{kpc}}^{-3}}\right).
   \label{ISM_crit_torrey_pap3.eqn}
  \end{equation}
Throughout this work, we use this dust-distribution recipe for the Auriga galaxies, this is the same as \texttt{recT12} in \citetalias{2021MNRAS.506.5703K}.

The diffuse dust in our simulations incorporates the \texttt{THEMIS} dust model, as described by \citet{2017A&A...602A..46J}. This model encompasses two types of dust grains: amorphous silicates and amorphous hydrocarbons. Regarding the amorphous silicates, it is assumed that half of the mass consists of amorphous enstatite, while the other half is composed of amorphous forsterite. The size distribution follows a lognormal distribution, which remains consistent for both populations of amorphous silicates, spanning sizes within the range of $a \simeq 10-3000~\mathrm{nm}$. The distribution peak occurs at $a_{\text{peak}} \simeq 140~{\text{nm}}$. For the amorphous hydrocarbon population, the size distribution comprises a combination of a power-law and a lognormal distribution. The power-law distribution is applied to amorphous carbon particles with sizes $a \lesssim 20~{\text{nm}}$, while the lognormal distribution is employed for larger grains within the size range of $a \simeq 10-3000~{\text{nm}}$. The lognormal distribution exhibits a peak at $a_{\text{peak}} \simeq 160~{\text{nm}}$. In this work, we used 15 bins to discretize the various grain size distributions.

\begin{table}
\caption{Input parameters of the SKIRT radiative transfer model for each of the Auriga components, in addition to the particle and cell positions.}
\label{Tab:Input model parameters}
\begin{tabulary}{\columnwidth}{LLL}
\hline
 Parameter \hspace{15mm} & Description & Origin \\
 & \vspace{.1cm} \centering \mbox{Dust} & \\
\hline
$\rho_\mathrm{gas}$ & Gas density & Simulation \\
$Z$ & Gas metallicity & Simulation \\
$T$ & Gas temperature & Simulation \\
$f_{\text{dust }}$ & Dust to metal fraction & Free parameter \\
 
\\
 & \centering \mbox{Old stellar population: BC03} & \\
\hline
$h$ & Smoothing length & Assumed distribution \\
$M_{\text {init }}$ & Initial mass & Simulation \\
$Z$ & Metallicity & Simulation \\
$t$ & Age & Simulation \\

\\
 & \centering \mbox{Star-forming regions: \texttt{TODDLERS}} & \\
\hline
$h$ & Smoothing length & Calculated \\
$M_{\text {init}}$ & Initial mass & Simulation \\
$t$ & Age & Simulation \\
$Z$ & Metallicity & Simulation \\
$\epsilon_{\text{SF}}$ & Star-formation efficiency & Free parameter \\
$n_{\text{cl}}$ & Density of natal clouds & Free parameter\\
$M_{\text{cl}}$ & Mass of natal clouds & Assumed power-law\\
$w$ & Normalization & Calculated \\
\hline
\end{tabulary}
\end{table}

\subsection{Star-forming regions}

The sub-resolution gas and dust emission is dealt with by assigning the young stellar particles in the simulation SEDs from the \texttt{TODDLERS} library. We earmarked all particles in the simulation snapshots with an age below $30~{\text{Myr}}$ as star-forming regions.

The \texttt{TODDLERS} library has five parameters, namely: the age of the system, $Z$, $\epsilon_{\rm{SF}}$, $n_{\rm{cl}}$, and $M_{\rm{cl}}$. The system's age corresponds to the overall evolution time and matches the age of the oldest cluster in the system, though younger stellar clusters may also be present due to recollapse events (see Sect.~{\ref{subsect:TODDLERS}}). The age and $Z$ were taken directly from the simulation and a power-law distribution with an exponent $-1.8$ is assumed for $M_{\rm{cl}}$. We treated $\epsilon_{\rm{SF}}$ and $n_{\rm{cl}}$ as free parameters to be calibrated. We assigned a single value to these parameters for the entire Auriga galaxy sample, although varying these values is straightforward. The calibration process for $\epsilon_{\rm{SF}}$ and $n_{\rm{cl}}$ is described in Sect.~{\ref{subsect:calibration}}. Finally, in addition to assigning \texttt{TODDLERS} model parameters to the star-forming particles, we normalized the assigned SED. This is described in the next subsection.
\begin{figure}
    \centering
\includegraphics[width=.9\columnwidth]{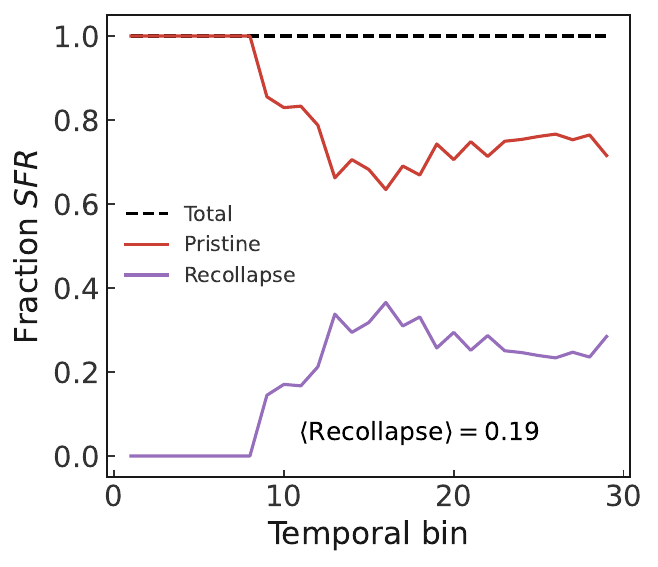}
    \caption{Example of the fractional SFR contribution originating from recollapsing shells using the method discussed in Sect.~\ref{sec:recollapse_handling_weights}. The system is assumed to have a uniform SFH and has \texttt{TODDLERS} parameters fixed at $Z=0.02,~ \epsilon_{\rm{SF}}=0.025,~ n_{\rm{cl}}=320\,\rm{cm^{-3}}$. The mean value of recollapsing shells is $19\,\%$.}
\label{fig:recollaspe_Contribution_Z02}
\end{figure}

\subsection{Matching model SFR with the simulation}
\label{sec:recollapse_handling_weights}

\begin{figure*}
\centering
\includegraphics[width=.85\textwidth]{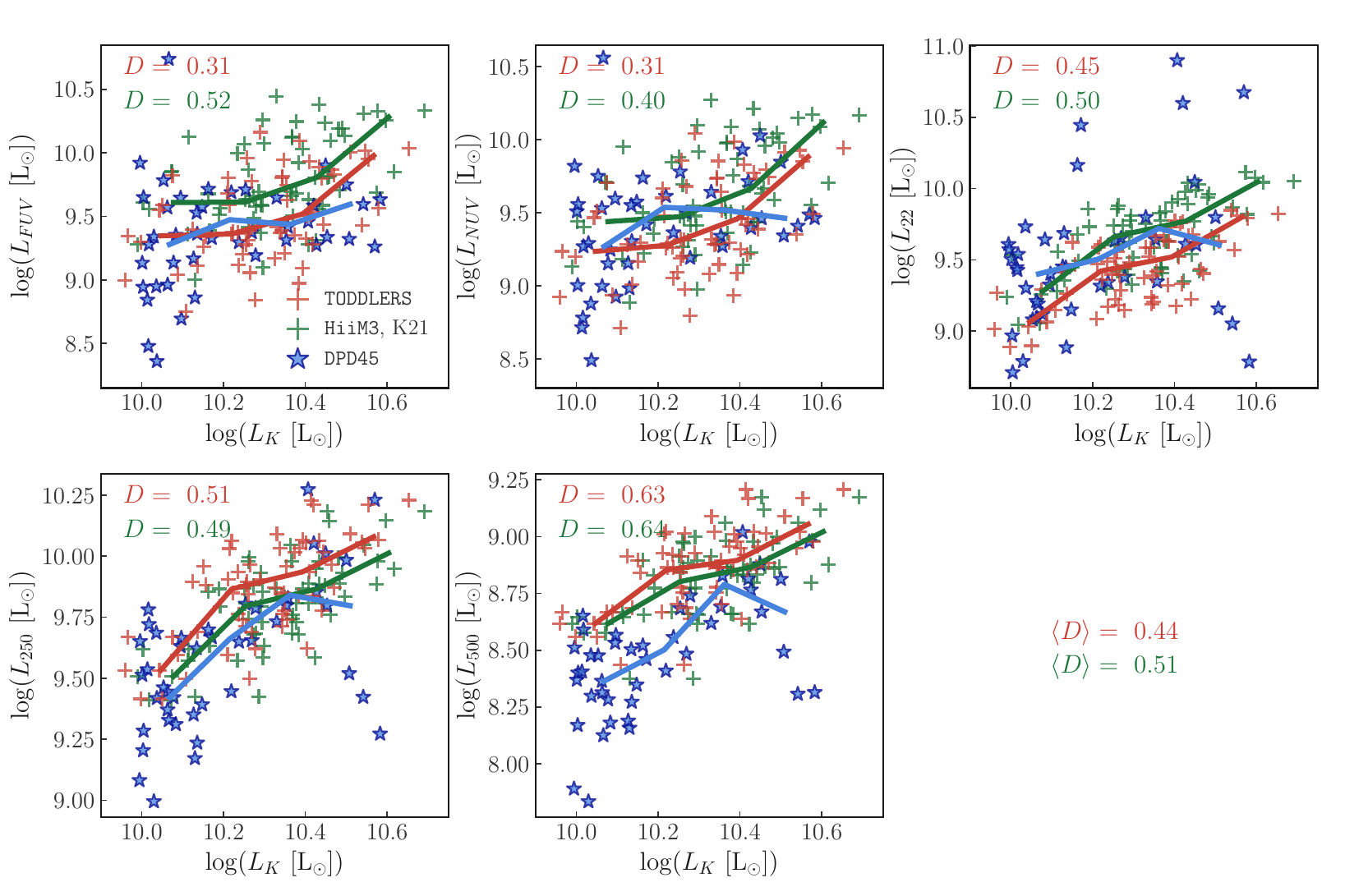}
\caption[Comparison of broad-band luminosity scaling relations between the DustPedia sub-sample and \texttt{SKIRT} post-processed Auriga galaxies using the two available SF regions libraries.) and \texttt{HiiM3} (green crosses) ]{Comparison of broad-band luminosity scaling relations between the DustPedia sub-sample \texttt{DPD44} (blue stars) and \texttt{SKIRT} post-processed Auriga galaxies using the two available star-forming regions' libraries in \texttt{SKIRT}, \texttt{TODDLERS} (red crosses), and \texttt{HiiM3} (green crosses) at the calibrated parameter values. The Auriga sample in both cases consists of 60 points, including an edge-on and face-on configuration for each of the 30 Auriga galaxies. The rolling median lines use four bins of equal width in the K-band luminosity, $L_{K}$.}
\label{fig:calib_scaling}
\end{figure*}

In the \texttt{TODDLERS} model, star-forming regions are capable of undergoing multiple episodes of star formation. In \citetalias{2023MNRAS.526.3871K}, we normalized the observables by the stellar mass at a given time in well-sampled cloud mass distributions parametrized by $Z$, $\epsilon_{\rm{SF}}$, and $n_{\rm{cl}}$. This presents a problem in the case of clouds exhibiting recollapse events due to an inconsistency in the stellar age. Indeed, in the case of recollapsing models, the overall age of the cloud is not representative of all the stellar generations present in it. 

To ensure alignment between the \texttt{TODDLERS} model's SFR and the SFR of the simulated galaxy, we divided the simulation star particles into sub-particles. Each sub-particle corresponds to one of the specific cloud masses defined in the \texttt{TODDLERS} model while having the other model parameters remain the same\footnote{$10^{5}-10^{6.75}\, \rm{M_{\odot}}$ in steps of 0.25 dex}. Each of the sub-particles was also assigned a weight, which can be modified to allow for the correct representation of the SFR despite the presence of recollapsing shells.

The calculation of the weights of the sub-particles consists of the following steps:
\begin{enumerate}[leftmargin=*]
\item As mentioned earlier, star particles younger than 30 Myr are tagged as star forming. The 30 Myr period is segmented into $N$ bins; here, we use $N=30$.
\item The initial weight for the $i^{\rm{th}}$ sub-particle, which corresponds to one of the eight cloud mass values in our parameter space, for the $j^{\rm{th}}$ star-particle is calculated as follows:
\begin{equation}
    w^{\rm{init.}}_{ij} = \frac{\chi^{i} \times M_{\rm{init}}^{j}}{\epsilon_{\rm{SF}} \times M_{\rm{cl}}^{i}},
    \label{eqn:initial_weights}
\end{equation}
where $\chi_{i}$ represents the fractional mass in each of the clouds and follows directly from the power-law mass distribution with the same exponent as \citetalias{2023MNRAS.526.3871K} ($\alpha = -1.8 $). $M_{\rm{init}}$ is the initial particle mass reported by the simulation.
We note that the division of particles into sub-particles is needed as the recollapse times are cloud mass-dependent (see \citetalias{2023MNRAS.526.3871K}), which are required for updating the initial weights.
\item To update the weights of sub-particles belonging to a given time bin, we systematically evaluate the impact of recollapsing models. Starting with the earliest time bin, we examine all bins prior to the given bin for the presence of recollapse events that could contribute stellar mass to the bin in question.
Given the simulation stellar mass in the $k^{\text{th}}$ time-bin, $M^{k}_{*}$ (or $\sum^{}_{j \in k} M_{\rm{init}}^{j}$) and the total stellar mass contributed by recollapsing shells from bins prior to that bin, $M^{k,\, \text{recoll.}}_{*}$,
we update the weight of sub-particle $i$ of the particle $j$ in time bin $k$ as: 
\begin{equation}
    w_{ij} = g_{k} \times w^{\rm{init.}}_{ij} ~ \forall ~ j \in k,
\text{ where }
    g_{k} = \frac{M^{k}_{*} - M^{k,\, \rm{recoll.}}_{*}}{M^{k}_{*}}.
    \label{eqn:final_weights}
\end{equation} 
\end{enumerate}

This approach effectively reflects the proportion of stellar mass not attributable to recollapsed shells. It ensures that the total stellar mass in a bin, which could include contributions from star formation in both clouds undergoing their first burst (pristine) and older systems showing recollapse, matches the total stellar mass given in the simulation. The weights calculated in Eq.~\eqref{eqn:final_weights} serve as normalization factors for the observables.

An example of the fractional SFR contribution originating from recollapsing shells using the above method is shown in Fig.~\ref{fig:recollaspe_Contribution_Z02}. The system is assumed to have a constant SFR and has \texttt{TODDLERS} parameters fixed at $Z=0.02$, $\epsilon_{\text{SF}} = 0.025$, and  $n_{\text{cl}} = 320~{\text{cm}}^{-3}$. In this case, star formation in recollapsing shells contributes approximately 20\% on average over the 30~Myr period.

It is worth mentioning that this method could lead to negative weights in some cases. An example of this could be the case of a system exhibiting a sharply declining star formation history (SFH) in the 30 Myr period. In these cases, ($\epsilon_{\rm{SF}}$, $n_{\rm{cl}}$) combinations which do not exhibit recollapse could be used instead.


\subsection{Older stellar population}
All stellar particles with an age greater than 30 Myr are assumed to be in systems that have cleared dust and gas around them. These are assigned SEDs from the \citet{2003MNRAS.344.1000B} template
library based on metallicity and age. We employ a Chabrier IMF while using this library.
Table~\ref{Tab:Input model parameters} summarizes the various parameters related to the input radiation sources and the dusty media, the last column states the source of the values used in the \texttt{SKIRT} input model.

\subsection{Radiative transfer configuration}

The \texttt{SKIRT} simulations carried in this work are broadly comprised of two steps:\ first, the propagation of  light originating from primary sources through the computational domain to assess the absorption of energy by interstellar dust; second, the propagation in the previous determines the temperature distribution, and thus, the thermal emission properties of the dust on a cell-by-cell basis within the domain. If the secondary emission is turned on, the thermal emission spectrum of each cell is propagated to generate mock observations at longer wavelengths.
Given the aforementioned procedure, the fidelity of a \texttt{SKIRT} simulation critically depends on the optimal discretization of the computational domain. To generate optimal spatial grids for Auriga galaxies, we used an octree grid with levels of 6 to 12 and a maximum cell dust fraction value of $10^{-6}$.

Regarding the wavelength grids employed in this work, we note that \texttt{SKIRT} employs separate, independent wavelength grids for the sources, the mean radiation field, the dust emission, and the instruments \citep{2020A&C....3100381C}. We use a logarithmic wavelength grid with 40 points running from $0.02\,\mu \rm{m}$ to  $10\,\mu \rm{m}$ for storing the mean radiation field in each cell. For dust emission, a nested logarithmic grid is employed. The low-resolution part of this nested grid has 100 points, running from $1\,\mu \rm{m}$ to  $2000\,\mu \rm{m}$, whereas the higher-resolution part runs in the polycyclic aromatic hydrocarbon (PAH) emission range from $2\,\mu \rm{m}$ to  $25\,\mu \rm{m}$ with 400 wavelength points.
\subsection{Calibration}\label{subsect:calibration}
Following \citetalias{2021MNRAS.506.5703K}, we used an SFR, stellar-mass, and morphology-based criteria to make a cut in the DustPedia sample. For the DustPedia galaxies, the SFR is the one obtained as a result of SED fitting carried out using the \texttt{CIGALE} code \citep{2012ASPC..461..569R, 2019A&A...622A.103B}. This SED fitting follows the settings described in \citet{2018A&A...620A.112B, 2019A&A...624A..80N, 2020MNRAS.494.2823T}. 
As the entire Auriga sample is star-forming at $z=0$, we consider DustPedia star-forming galaxies with ${\mathrm{SFR}}_{\mathrm{cig}}>~0.8~\mathrm{\rm{M_{\odot}}yr^{-1}}$, where ${\mathrm{SFR}}_{\mathrm{cig}}$ is the SFR inferred using the \texttt{CIGALE} SED fitting code.
The stellar mass is tracked by the flux in the 2MASS-K band, with a pivot wavelength of $2.16\,\mu \rm{m}$. We note that this is different from \citetalias{2021MNRAS.506.5703K}, where the WISE-W1 band was used as a proxy for the stellar mass. We make this choice in order to avoid the contribution of the PAH features that contribute to this band. We chose all the DustPedia galaxies that fall within the K-band luminosity range of the Auriga sample.
Finally, the DustPedia dataset is limited to late-type galaxies (Hubble stage > 0 in the DustPedia catalog). These cuts in the Dustpedia sample result in 45 galaxies, which we call $\texttt{DPD45}$ in the text. 

\begin{figure*}
\centering
\includegraphics[width=.75\textwidth]{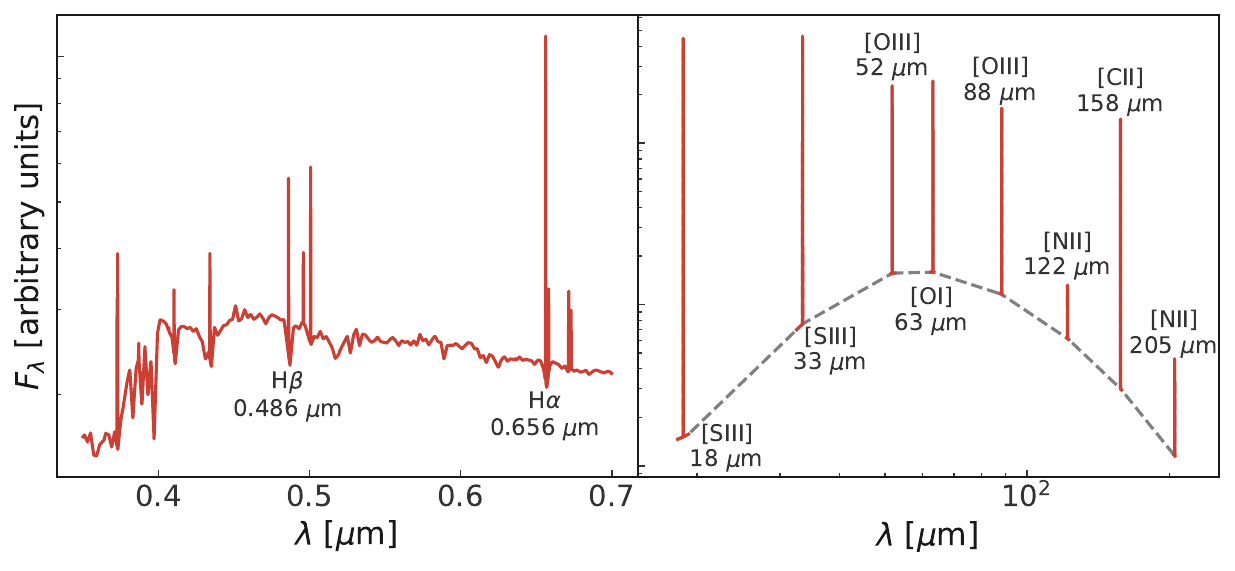}
\caption[Observables generated for the line-emission-based analysis carried out in this work.]{Observables generated for the line-emission-based analysis carried out in this work. Left: Spectrum at a resolution $R=3000$ in the optical $0.4-0.7\,\mu \rm{m}$, right: Selected windows for lines in the FIR, again at $R=3000$. At each of these wavelengths, we also generate images at a spatial resolution of $0.5\,\rm{kpc}$}
\label{fig:IFU_spectra_example}
\end{figure*}

We keep the dust-to-metal fraction for the diffuse dust fixed at the value found in \citetalias{2021MNRAS.506.5703K}, namely, $f_{\text{dust}}=0.14$. Keeping this value fixed ensures that a consistent comparison can be made with the results obtained in that paper, where \texttt{HiiM3} was employed as the emission library for star-forming regions.
On the other hand, the free parameters of the \texttt{TODDLERS} model ($\epsilon_{\text{SF}}$, $n_{\text{cl}}$) are found by comparing broadband luminosity scaling relations between the post-processed Auriga galaxies and \texttt{DPD45}.
To quantify the effect of changing the free parameters, we used a two-dimensional (2D) generalization of the well-known Kolmogorov–Smirnov (K-S) test \citep{1983MNRAS.202..615P}. The K–S test computes a metric $D$ which can be interpreted as a measure of the "distance" between two sets of 2D data points, with smaller $D$ values indicating better correspondence.
The selected values of the free parameters are the ones that minimize the average $D$ value obtained by applying the K-S test on set of scaling relations. We chose the scaling relation between 2MASS-K, and each one of GALEX-FUV, GALEX-NUV, WISE-22, SPIRE-250, and SPIRE-500 for this purpose. We note that our calibration is robust regardless of the method used to calculate the distance between \texttt{DPD45} and the synthetic datasets. We confirmed this by testing with the 2D Wasserstein distance \citep{2015JMIV...51...22B}.

By scanning the \texttt{TODDLERS}' parameter space, we find that the parameter set $\epsilon_{\text{SF}} = 0.025$, $n_{\text{cl}}=320 ~\mathrm{cm^{-3}}$ gives a very good match to the observational data.
We note that while the minimum of 
the average $D$ is degenerate to the free parameters of the model, the chosen parameter set shows a good agreement across all the broadbands used for calibration.
Figure~\ref{fig:calib_scaling} shows the scaling relations when the optimal values of the free parameters are employed. 
Also shown in Fig.~\ref{fig:calib_scaling} are the scaling relations obtained for the Auriga galaxies using \texttt{HiiM3} star-forming regions' library and the calibration described in \citetalias{2021MNRAS.506.5703K}.

The scaling relations obtained with the calibrated \texttt{TODDLERS} parameter set exhibit a higher level of consistency with the observational data when compared to the \texttt{HiiM3} calibrated scaling relations, as indicated by the average $D$ value. It is especially worth mentioning the enhanced attenuation in the GALEX bands resulting from the change in the star-forming regions' library,  where the average decrease in the FUV band with respect to the runs in \citetalias{2021MNRAS.506.5703K} is  $\approx0.3$ dex. This is an encouraging development in light of the findings in previous studies of a similar nature which found higher FUV and NUV luminosities when compared to observations. These discrepancies were attributed to the star-forming regions library \citep[e.g.,][]{2019MNRAS.484.4069B, 2020MNRAS.494.2823T, 2021MNRAS.506.5703K, 2022MNRAS.512.2728C, 2022MNRAS.516.3728T,2024MNRAS.531.3839G}.

\section{Synthetic data products}
\label{sect:data_products}

For the purpose of this work, we generated the following synthetic data products for the Auriga galaxies:
\begin{enumerate}[leftmargin=*]
    \item Broadband images were generated in the same set of 50 broadbands as those used in \citetalias{2021MNRAS.506.5703K}. Additionally, we generated transparent images in the UV and optical bands. To achieve this, we used the incident stellar spectra from \texttt{TODDLERS} models and removed the diffuse dust.
    \item High spectral resolution images were generated in the wavelength range of $0.3-0.7\,\mu \rm{m}$ with a spectral resolution of $R=\lambda/\Delta \lambda=3000$. We used \texttt{SKIRT} in the extinction-only mode for this set. An example of the result obtained by spatially integrating such images is shown on the left side of Fig.~\ref{fig:IFU_spectra_example}. Apart from this, we also calculate transparent images without diffuse dust and foreground dust attenuation in the \texttt{TODDLERS} models. However, the \texttt{TODDLERS} models do consider the Lyman continuum (LyC) absorption by dust. We note that the choice of the spectral resolution for the images is not limited by \texttt{TODDLERS} library, which was sampled at $R=5\times 10^{4}$. This choice is simply motivated by the computation time and the fact that we did not consider any kinematics in this work. 
    \item High spectral resolution images were generated in the FIR wavelength range of $10-200\,\mu \rm{m}$. We focused on small windows within this wavelength range to generate luminosity maps of a set of fine-structure lines. The used lines and the windows are shown on the right-side panel of Fig.~\ref{fig:IFU_spectra_example}. 
    We note that these images are the projections of light from the star-forming regions alone and do not take into account any diffuse media.
    \item Projected physical property maps: These maps include 2D projections of the physical properties such as the SFR surface density averaged over a certain period, the stellar mass surface density ($\Sigma_{\star}$), mass-weighted metallicity, and mass-weighted stellar ages. These maps serve as the true values whenever we carried out a a comparison with synthetic observations.\end{enumerate}

The images and the projected physical property maps were generated with a spatial resolution $\Delta x= 500~{\text{pc/pix}}$, and at four inclinations: 0$^\circ$, 30$^\circ$, 60$^\circ$, and 90$^\circ$. The field of view (FOV) was fixed to 90 kpc for all maps.
The optimal choice of the number of photon packets for the \texttt{SKIRT} runs is dependent on both the spatial and spectral resolution. The choices made in this work are discussed in Appendix \ref{appendix:choice of photon packets}.

\section{UV-mm SEDs and broadband maps}
\label{sec:UV-mm SED and broadband data analysis}

\begin{figure*}
\centering
\includegraphics[width=.85\textwidth]{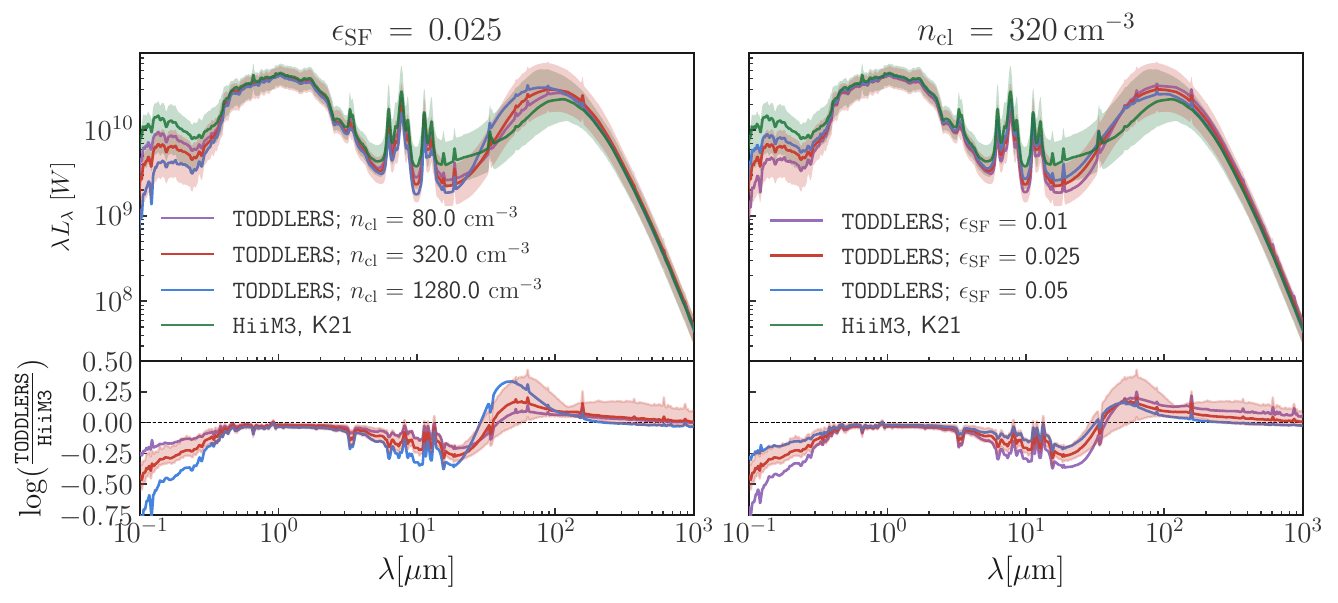}
\caption{Comparison of the median UV-mm SEDs from the \texttt{SKIRT} post-processed Auriga galaxies using the two available star-forming regions libraries in \texttt{SKIRT}, \texttt{TODDLERS} (red) and \texttt{HiiM3} (green) at the calibrated parameter values. The median is calculated by considering only the exactly edge-on and face-on inclinations. The corresponding filled areas give the range of values encountered for the entire Auriga sample. The lower panels show the ratio of the median SED obtained using \texttt{TODDLERS} and \texttt{HiiM3} libraries, with the range shown by the filled area only for the fiducial run.}\label{fig:calib_SED}
\end{figure*}

In this section, we focus on the global Auriga SEDs and the broadband maps generated from our post-processing procedure. The SEDs provide a convenient way to highlight the differences arising from variations in sub-grid treatment and calibration choices. Following this, we use the broadband maps to produce maps of quantities related to the correction of dust-obscured UV. Additionally, we examine resolved maps of MIR-FIR colors and compare them to observational data.

\subsection{UV--mm SEDs: \texttt{TODDLERS} vs. \texttt{HiiM3}}
\label{sect:GlobalSEDs_impact_of_subgrid_treatment}

In Sect.~\ref{subsect:calibration} we showed that it is possible to successfully calibrate the \texttt{TODDLERS} parameters to reproduce observational scaling relations. In this section we consider the Auriga UV-mm SEDs and the implications of changing the of star-forming regions' sub-grid treatment as well as varying the \texttt{TODDLERS} parameters around the calibrated values.
Figure~\ref{fig:calib_SED} shows the median SEDs for all the Auriga galaxies for various combinations of the two \texttt{TODDLERS} parameters along with the median SED obtained in \citetalias{2021MNRAS.506.5703K} using the \texttt{HiiM3} library. The filled area gives the range of values encountered for the entire Auriga sample. The median values and the ranges are calculated using SEDs at $0^{\circ},\,90^{\circ}$ for each of the Auriga galaxies.

The left panel in Fig.~\ref{fig:calib_SED} shows the impact of changing $n_{\rm{cl}}$ in the range $80-1280\, \rm{cm^{-3}}$ while keeping fixed at $\epsilon_{\rm{SF}}=0.025$, while in the right panel $\epsilon_{\rm{SF}}$ is varied from $0.01-0.05$ while keeping $n_{\rm{cl}}$ fixed at $320\,\rm{cm^{-3}}$. In comparison to the \texttt{HiiM3} library, with its free parameters calibrated in \citetalias{2021MNRAS.506.5703K}, the most significant differences are in the UV attenuation, MIR emission, and the FIR emission in the 50-100$\,\mu \rm{m}$ spectral regimes. A comparison of the images obtained using these two libraries in selected bands is shown in Fig.~\ref{fig:HiiM3_vs_TODD_images}.
The dust emission characteristics are primarily a result of the non-PAH dust size distribution in \texttt{TODDLERS} having a cut-off at $0.03\,\mu \rm{m}$. This implies that the MIR emission contribution from non-PAH dust is only sizeable when the radiation intensity is high, namely, during the early stages of evolution. Instead, the grain size distribution leads to a stronger emission in the 50-100$\,\mu \rm{m}$ range. 

We note that the changes in UV attenuation and MIR emission, resulting from the altered sub-grid treatment of emission from star-forming regions, reduce the discrepancies with observational data reported in earlier studies of similar nature (see Sect.~\ref{subsect:calibration}). Additionally, a recent comparison of IllustrisTNG-100 galaxies post-processed with \texttt{HiiM3} templates has shown that these galaxies produce bluer FUV-NUV colors compared to observational data \citep[][]{2024MNRAS.531.3839G}. By changing the star-forming regions' treatment to \texttt{TODDLERS}, this discrepancy is reduced. This can also be inferred from the redder UV slopes in the SEDs produced using \texttt{TODDLERS} in Fig.~\ref{fig:calib_SED}
, compared to those produced with \texttt{HiiM3}. The better agreement with observational data is not limited to the UV and UV colors; in Sect.~\ref{Sec:auriga_MIR-FIR colors}  we demonstrate this for the resolved MIR-FIR colors of the Auriga sample.

As for the impact of changing the two parameters of \texttt{TODDLERS}, increasing $n_{\rm{cl}}$ at a fixed $\epsilon_{\rm{SF}}$ leads to more compact star-forming regions. This tends to increase the UV attenuation in Auriga galaxies and a shift of the FIR peak to lower wavelengths. On the other hand, increasing $\epsilon_{\rm{SF}}$ at a fixed $n_{\rm{cl}}$ leads to a lower UV attenuation, with a slight movement of the FIR peak to lower wavelengths.
It is worth noting that the attenuation in the star-forming regions also impacts the emission by the diffuse dust, which is not deficient in small grains. In particular, the single-photon heating contribution in the MIR goes up with \texttt{TODDLERS} parameters which promote a rapid dissolution of the clouds, namely, at lower $n_{\rm{cl}}$ and/or higher $\epsilon_{\rm{SF}}$. Rapid shell dissolution ensures that the UV from star-forming regions escapes with a lower attenuation and heats the diffuse dust.

\subsection{UV-mm broadband maps}
\label{sec:UV-mm broadband data analysis}

\begin{figure*}
\centering
\includegraphics[width=.8\textwidth]{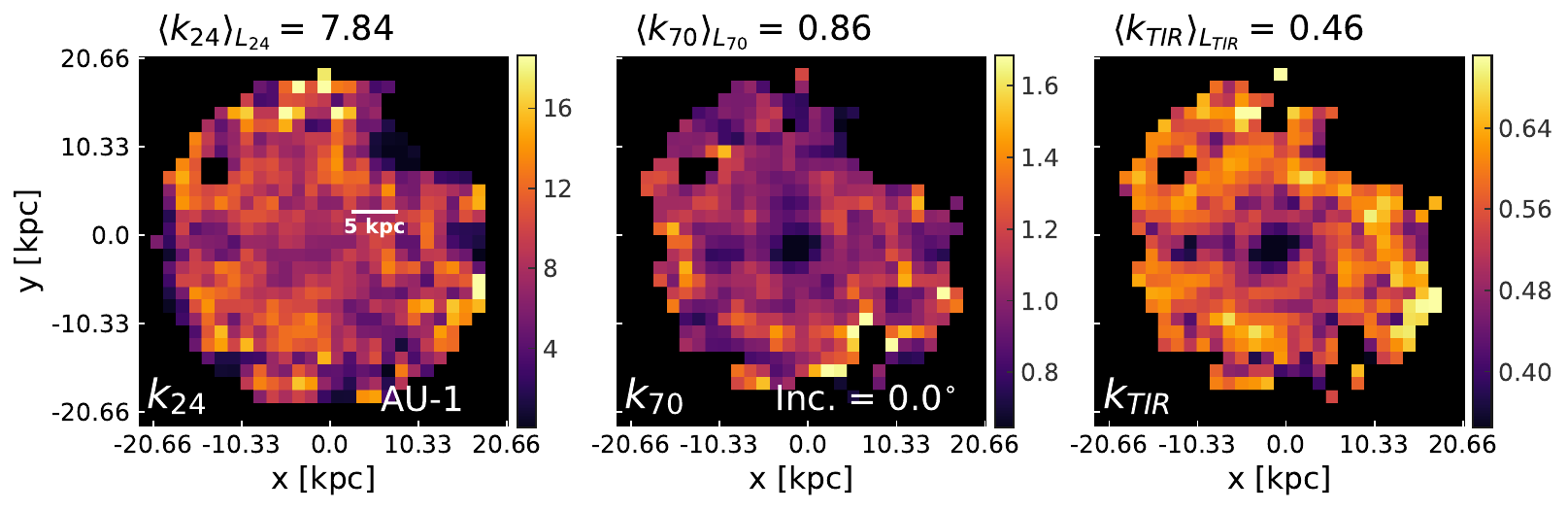}
\includegraphics[width=.8\textwidth]{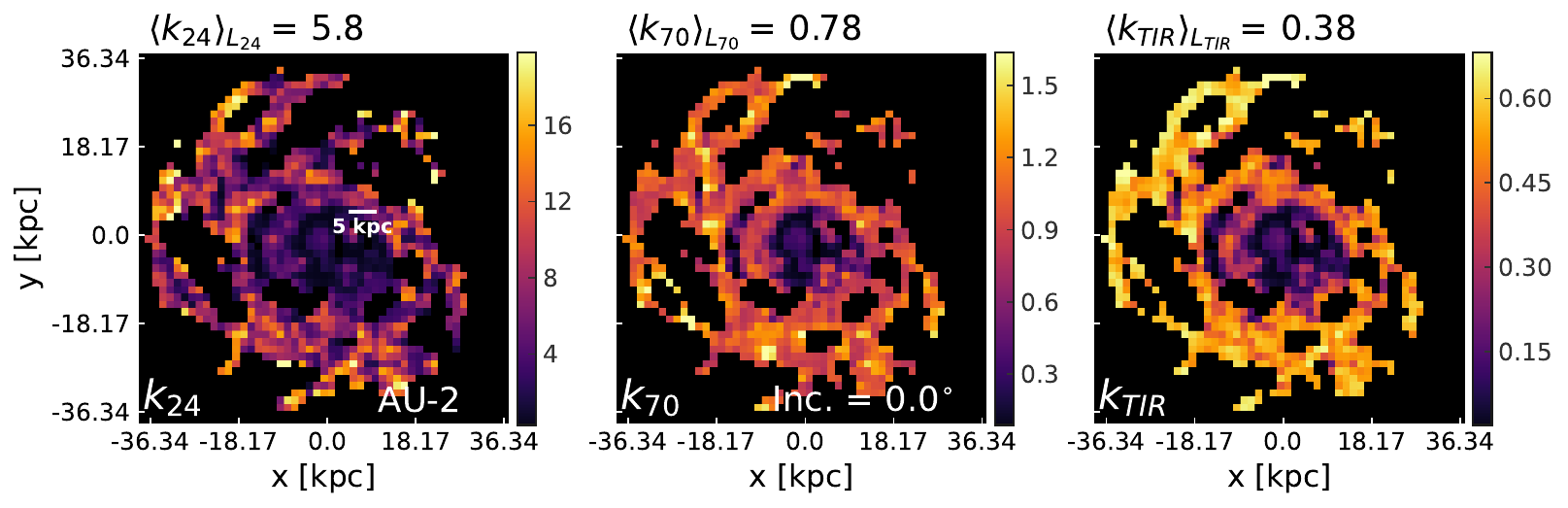}
\includegraphics[width=.8\textwidth]{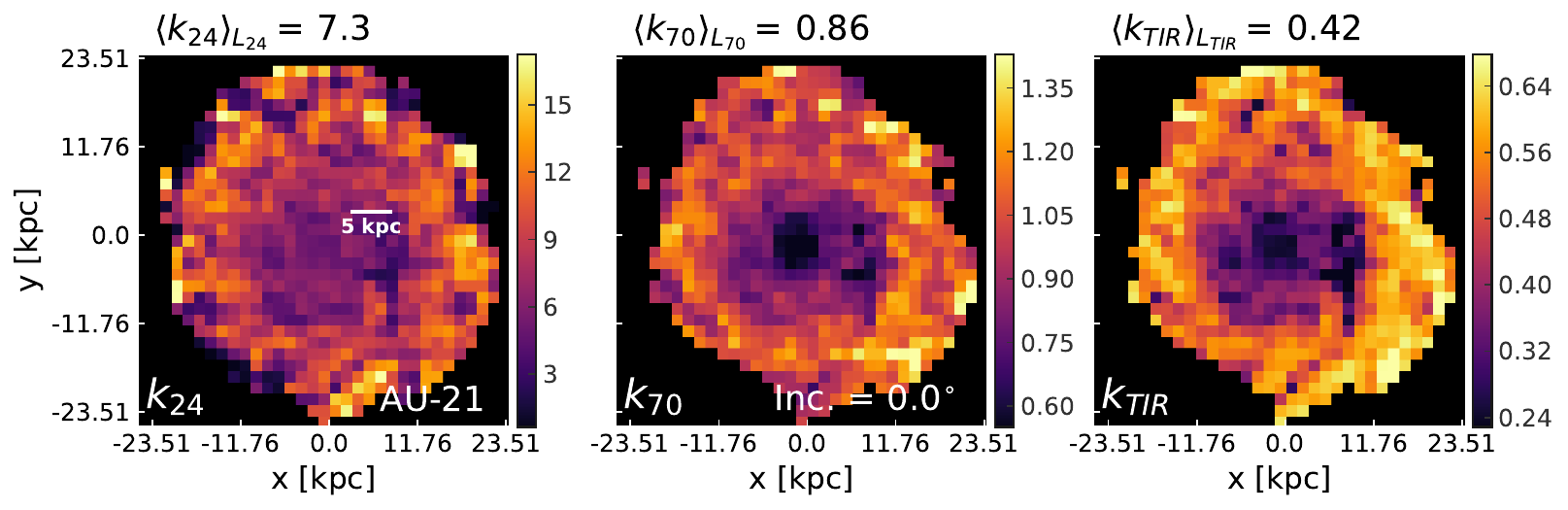}
\caption{Spatially varying IR correction factor, $k_{IR}$ for three Auriga galaxies in a face-on configuration. From left to right: $k_{24}$, $k_{70}$, and $k_{TIR}$ for AU-21 (top), AU-23 (middle), and AU-27 (bottom). Each panel also lists the luminosity weighted mean of the value obtained using all the pixels.}
\label{fig:spatially_varying_kIR}
\end{figure*}

\begin{figure}
    \centering
\includegraphics[width=.9\columnwidth]{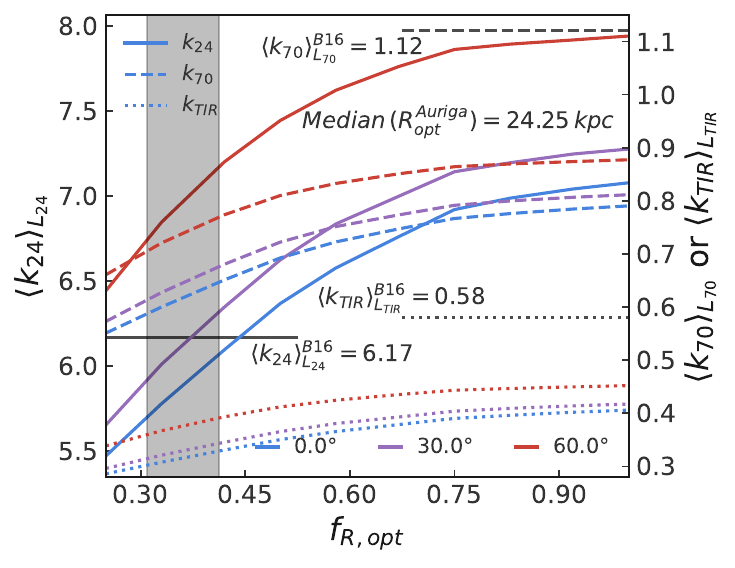}
    \caption{Mean light-weighted IR correction factors, $\left < k_{IR} \right >_{IR}$ as a function of the aperture and the inclination using the galaxies in the \texttt{Auriga-D} sample. The aperture is given as a fraction of the optical radius, with the gray region denoting a median aperture of 15-20 kpc of the simulated galaxies. The values from B16 are shown as gray lines.}
    \label{fig:k_variation_aperture}
\end{figure}

\begin{table*}
\caption{Comparison of $k_{\text{IR}}$ values obtained for the \texttt{Auriga-D} sample and the observational sample considered in this work.}
\hspace{-0.425cm} 
\begin{minipage}{\textwidth}
\centering
\begin{tabular}{lcccccccc}
\toprule
Band & $\left<k_{IR}\right>^{f_{\text{R,opt}}=1}$ & $\left<k_{IR}\right>^{f_{\text{R,opt}}=1}_{L_{IR}}$ & $\left<k_{IR}\right>^{f_{\,\text{R,opt}}=0.33}$ & $\left<k_{IR}\right>^{f_{\,\text{R,opt}}=0.33}_{L_{IR}}$ & $\left< k_{IR}\right>$,\,B16 & $\left<k_{IR} \right>_{L_{IR}}$,\,B16 & $k_{IR}$,\,H11&  $k_{IR}$,\,L11 \\
\midrule
24 & $7.95 \pm 4.25$ & $7.43 \pm 2.83$ & $6.12 \pm 2.72$ & $6.25 \pm 1.91$ & $8.11 \pm 2.10$ & $6.17 \pm 2.17$ & $3.89 \pm 0.15$ & 6.0 \\
70 & $1.03 \pm 0.30$ & $0.83 \pm 0.30$ & $0.81 \pm 0.28$ & $0.65 \pm 0.25$ & $1.57 \pm 0.39$ & $1.12 \pm 0.50$ & - & - \\
TIR & $0.50 \pm 0.15$ & $0.43 \pm 0.15$ & $0.36 \pm 0.14$ & $0.33 \pm 0.11$ & $0.63 \pm 0.12$ & $0.58 \pm 0.12$ & $0.46 \pm 0.12$ & - \\
\bottomrule
\label{Table:correctionFactor_k_IR_broadband}
\end{tabular}
\tablefoot{
The first four columns show the mean $k_{\text{IR}}$ values for the \texttt{Auriga-D} sample, obtained across three bands and using three different face-on galaxy inclinations. The superscripts in these columns represent the aperture, while the subscripts, when present, indicate the luminosity weighting in the average.
Cols. 5 and 6 display corresponding results for the B16 dataset, while the final two columns report values from H11 and L11 studies.
}
\end{minipage}
\end{table*}

To carry out a spatially resolved analysis on the broadband images generated by our \texttt{TODDLERS-SKIRT} post-processing pipeline, 
we used all the pixels within $R_{\mathrm{opt}}$\footnote{$2 \times R_{\mathrm{opt}}$ is always smaller than the original FOV of 90 kpc used for our maps.}, the optical radius defined in \citet[][Table\,1]{2017MNRAS.467..179G}. 
We assumed the native physical resolution of our images (500 pc/pix) to be equivalent to an angular resolution of 12". All images were then convolved with a SPIRE-500$\,\mu\mathrm{m}$ point spread function (PSF) using the convolution kernels of \citet{2011PASP..123.1218A}. 
This was followed by a re-gridding procedure that generates a photometric grid across each galaxy containing individual squares 36"×36" in size. These squares have a side corresponding to a physical scale of 1.5 kpc, which is comparable to the physical pixel scale of the observational data we employ for comparison. The convolution process uses \texttt{astropy.convolution.convolve\textunderscore fft}\footnote{\url{https://docs.astropy.org/en/stable/api/astropy.convolution.convolve_fft.html}} and the re-gridding employs the \texttt{reproject} package\footnote{\url{https://reproject.readthedocs.io/en/stable/}} with \texttt{reproject\textunderscore exact} algorithm.
We apply the aforementioned data processing method to each of the Auriga galaxies, considering three face-on inclinations: $0.0^\circ$, $30.0^\circ$, and $60.0^\circ$.

To maintain consistency with the observational datasets used in this section, we apply a selection criterion to the Auriga galaxies based on their disc-mass-fraction ($f_{\text{disc}}$). We use the values reported in \citet[][Table~1]{2017MNRAS.467..179G} and select galaxies with $f_{\text{disc}}\geq 0.5$. This criterion results in a sub-sample of 22 galaxies, which we refer to as \texttt{Auriga-D}. This selection process effectively removes the most irregular and bulge-dominated systems, leaving us with a disk-dominated subset.

In the following discussion, the GALEX-$FUV$, IRAC\,$\,8\,\mu \rm{m}$, MIPS\,$\,24\,\mu \rm{m}$, PACS $\,70\,\mu \rm{m}$, PACS\,$\,100\,\mu \rm{m}$, and SPIRE\,$\,500\,\mu \rm{m}$ bands are simply referred to as $FUV$, 8, 24, 70, 100, and 500 $\mu \rm{m}$ bands, respectively.

\subsubsection{SFR using FUV and IR broadband data: Spatially varying IR correction factors}
\label{subsubsect:spatially_varying_kIR}
Given that star-forming regions are often dust obscured, accurately measuring SFR necessitates the tandem use of indicators which capture obscured and unobscured star formation. To that end, hybrid calibrations using FUV and IR broadband data are available in the literature, e.g., \citet[hereafter referred to as L11]{2011ApJ...735...63L} and those in \citet[hereafter referred to as H11]{2011ApJ...741..124H}. These can be written as:

\begin{equation}
\text{SFR}\left(\rm{M_{\odot}} \text{yr}^{-1}\right)=C_{1}\,[L_{FUV} + k_{IR}\cdot L_{IR}].
\label{eqn:SFR_UVIR}
\end{equation}

In Eq.~\eqref{eqn:SFR_UVIR}, $C_{1} = 4.6 \times 10^{-44}$ assuming a stellar IMF as given in \cite{2001MNRAS.322..231K}, with a constant SFR maintained over a period of 100 Myr, and using solar metallicity stellar models from \citet{1999ApJS..123....3L}. $k_{IR}$ depends on the IR band used for the correction. The L11 calibration using MIR gives $k_{24}=6.0$ , and the calibration in H11 with total-infrared (TIR) uses  $k_{TIR}=0.46$. The quantities within the square brackets in Eq.~\eqref{eqn:SFR_UVIR} indicates the FUV luminosities that have been corrected to their intrinsic values. The luminosities in these equations are measured in units of $\mathrm{erg\, s^{-1}}$. 

\begin{figure*}
    \centering
\includegraphics[width=.9\textwidth]{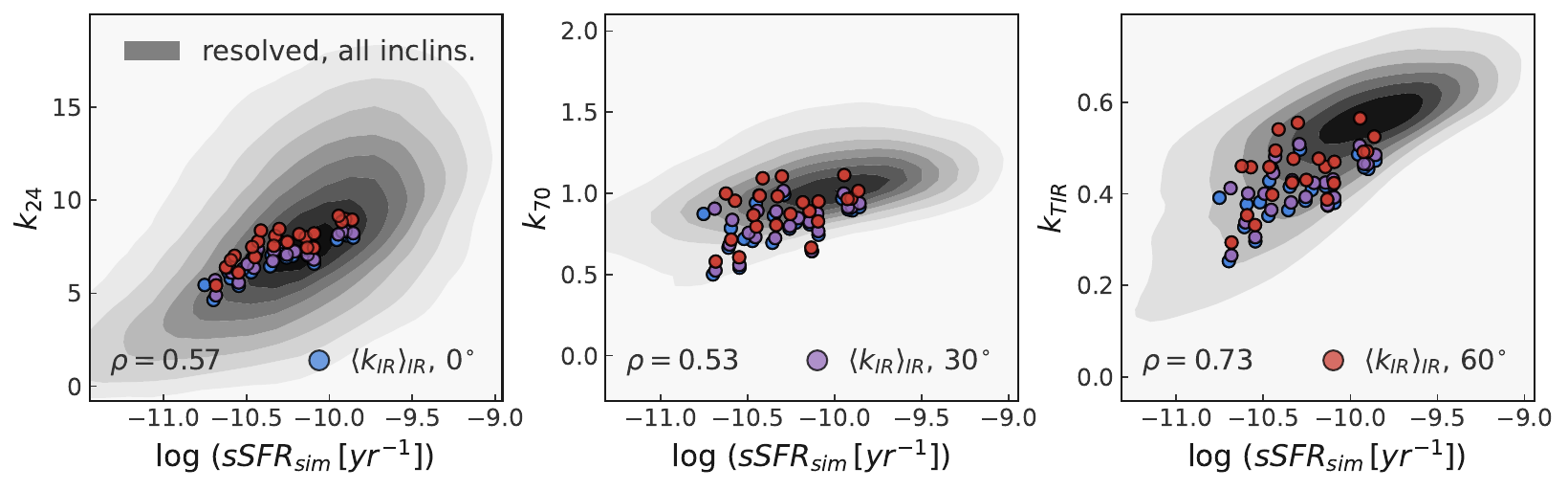}
    \caption{Spatially varying IR correction factor, $k_{IR}$, shown as a function of the projected sSFR for the \texttt{Auriga-D} sub-sample using the three face-on inclinations. 
    The circular points represent the light-weighted values for individual galaxies at the three inclinations considered in this work.}
\label{fig:k_vs_sSFR_aperture=1.0}
\end{figure*}

\begin{figure}
    \centering
\includegraphics[width=.85\columnwidth]{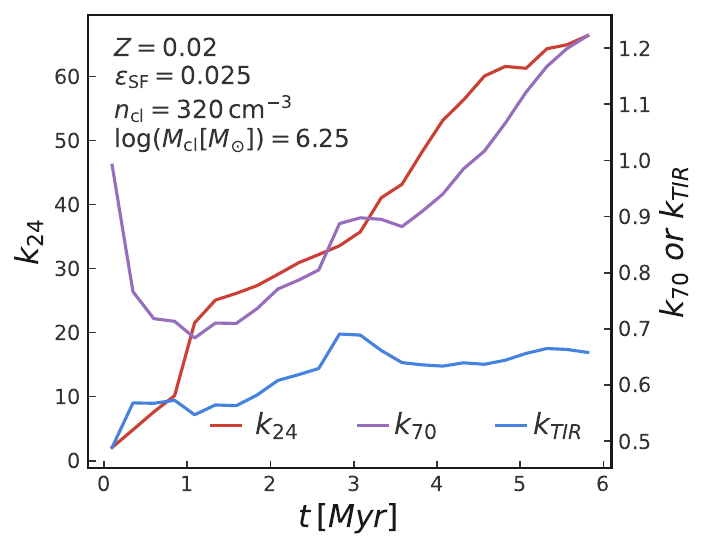}
    \caption{$k_{IR}$ as a function of time for an individual model from the \texttt{TODDLERS} library.}
    \label{fig:k_variation_individual_TODDLERS}
\end{figure}

\begin{figure*}
\centering
\includegraphics[width=.75\textwidth]{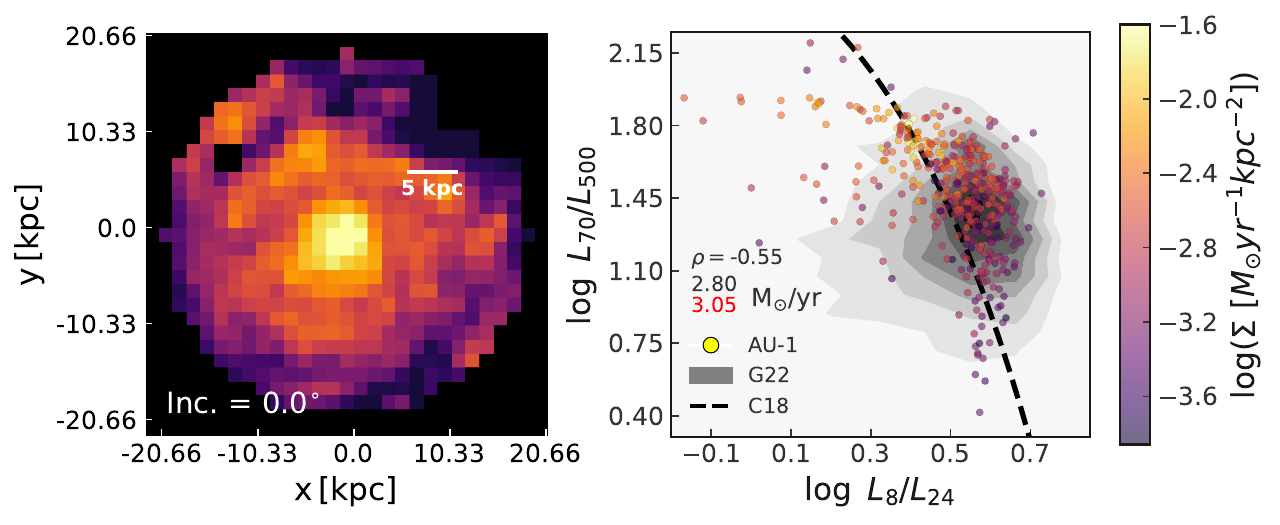}
\includegraphics[width=.75\textwidth]{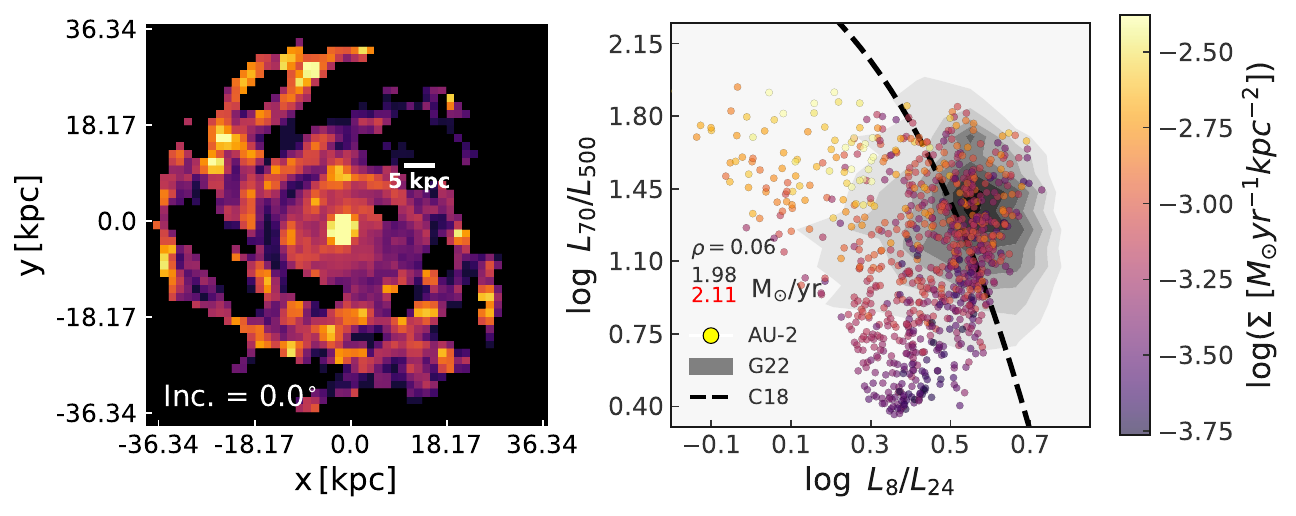}
\includegraphics[width=.75\textwidth]{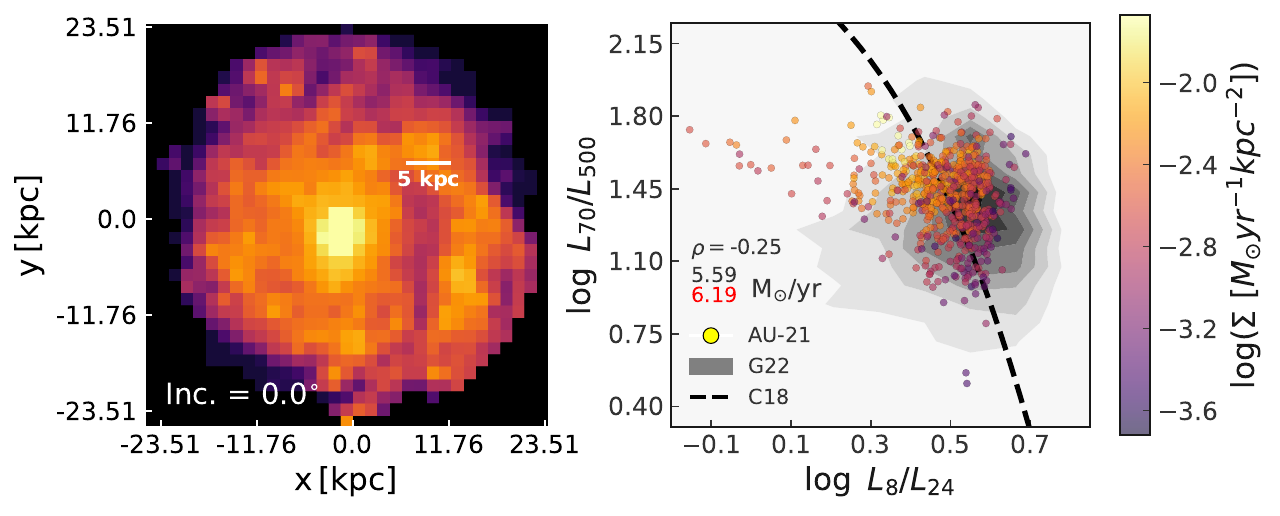}
\caption{SFRD maps (left) and associated MIR-FIR color-color relations (right) for three Auriga galaxies AU-1, AU-2, and AU-21 post-processed with \texttt{TODDLERS} models. The figures in the right column also indicate the PCC value, the SFR of the galaxy inferred using the calibration given in L11 (black value; see Sect.~\ref{subsubsect:spatially_varying_kIR}), and the true SFR of the galaxies averaged over 100 Myr (red value). The gray filled contours represent the colors of the KINGFISH sample as reported by G22 while the dashed curve is the relation for NGC-4449 given in C22.}
\label{fig:MIR_FIR_example}
\end{figure*}

\begin{figure*}
\centering
\includegraphics[width=.9\textwidth]{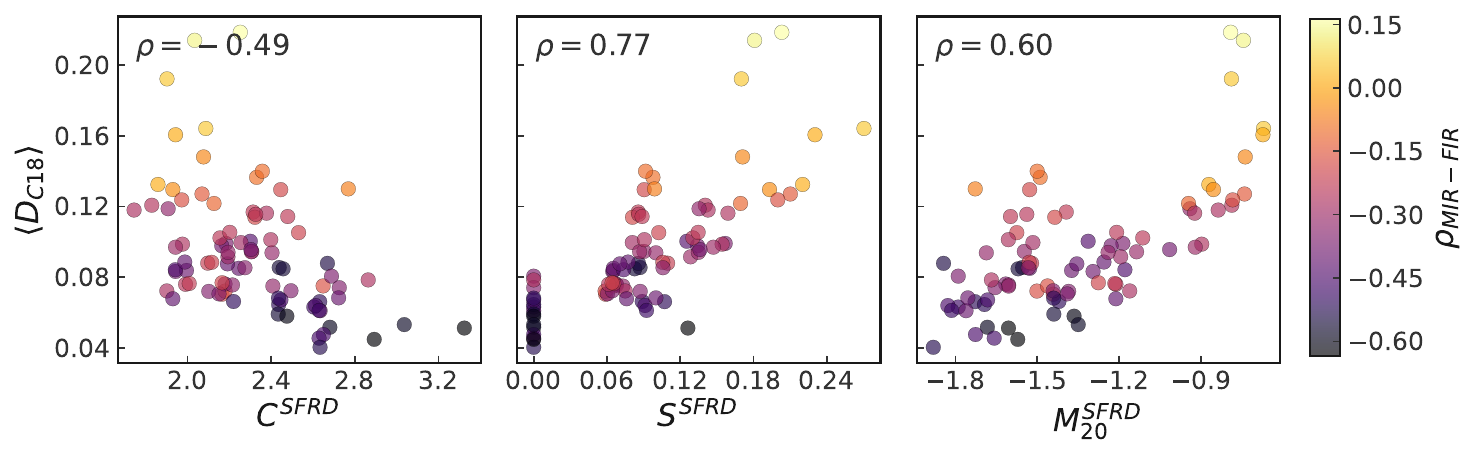}
\caption{Dispersion in Auriga galaxies' MIR-FIR color-color plot as a function of the non-parametric indicators of galaxy morphology applied to the SFRD maps. The data includes all the three inclinations and all Auriga galaxies except AU-11, which is undergoing a merger. The data-points are color-coded using the Pearson correlation coefficient obtained for the color-color relation for individual galaxies.}
\label{fig:Statmorph_SFRmaps}
\end{figure*}

\begin{figure}
\centering
\includegraphics[width=.85\columnwidth]{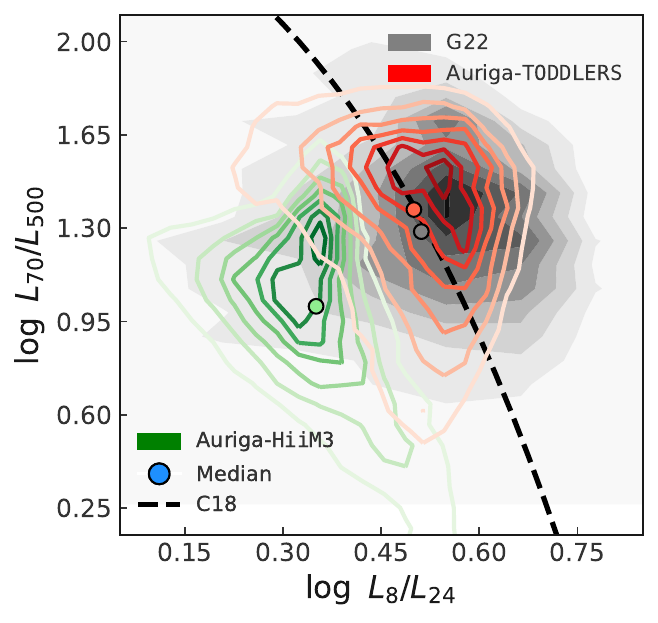}
\caption{MIR-FIR color-color diagram considering all the pixels of the \texttt{Auriga-D} galaxies at inclinations $0^{\circ},\,30^{\circ},\text{ and }60^{\circ}$ along with consolidated "Regime-1" and "Regime-2" data from G22 (gray region). The orange contours represent the data when \texttt{TODDLERS} library is employed, while the green contours are obtained in the case of \texttt{HiiM3}. The darkness of the contour colors represents the density of the pixels on a linear scale. Also shown are the median values for each of the three datasets using round markers of corresponding colors. The dashed black curve is the relation obtained in C18.}
\label{fig:MIRFIR_TODDvsHiiM3}
\end{figure}

Instead of employing a fixed $k_{IR}$ framework , as utilized in L11 and H11, 
local variations in $k_{IR}$ were recognized in \citet[hereafter referred to as B16]{2016A&A...591A...6B}  by making use of the SED-fitting driven spatially resolved estimation of the FUV attenuation.  Their results are based on a sample of eight star-forming spiral galaxies: NGC 628, NGC 925, NGC 1097, NGC 3351, NGC 4321, NGC 4736, NGC 5055, and NGC 7793. SED fitting was done using these galaxies, primarily of Sa type or later, at a kpc resolution.
The analysis in B16 ties $k_{IR}$ variability to specific galactic properties, such as the specific star-formation rate (sSFR), and invites a deeper investigation into the systematics affecting $k_{IR}$, specifically through the lens of simulated galaxies. The test-bed provided by simulated galaxies allows for an in-depth analysis of how factors like galaxy inclination and chosen apertures influence SFR estimations without assumptions made in SED fitting, like fixed galaxy metallicity. Furthermore, the intrinsic FUV luminosity on resolved scales is readily available. 
Given both the attenuated and unattenuated FUV values, we could find spatially varying infrared correction factors $(k_{IR}$). We solve the following equation to do so:
\begin{equation}
k_{IR} = \frac{L_{FUV,\,\rm{intr.}} - L_{FUV}}{L_{IR}}, \label{eqn:find_kIR_spatially_resolved}
\end{equation}
where $L_{FUV,\,\rm{intr.}}$, $L_{FUV}$, and $L_{IR}$ are the intrinsic GALEX-FUV due to the stellar particles (including the unattenuated UV from star-forming regions), the observed UV as given by the synthetic maps generated by \texttt{SKIRT}, and the values given by the IR synthetic maps, respectively. 
We use $24$, $70$, and TIR maps for the calculation of $k_{IR}$. The TIR is based on the three-band approximation given in Table~3 of \citet{2013MNRAS.431.1956G} which used $24$, $70$, and $100\,\mu \rm{m}$ bands to get the TIR emission. This approximation performs the best based on their goodness-of-fit criteria. 

Equation~\eqref{eqn:find_kIR_spatially_resolved}  applied to the post-processed Auriga galaxies probes the connection between the component of the radiation heating the dust largely attributable to star formation (numerator in Eq.~\eqref{eqn:find_kIR_spatially_resolved}) and the dust re-emission characteristics in realistic radiative transfer settings.
Figure~\ref{fig:spatially_varying_kIR} shows $k_{IR}$ in the case of $24\,\mu \rm{m},\,70\,\mu \rm{m}$, and TIR for three Auriga galaxies, AU-1, AU-2, and AU-21 for all pixels within the $R_{\rm{opt}}$. The local $k_{IR}$ values can vary significantly from the single values used in the literature (see Table~\ref{Table:correctionFactor_k_IR_broadband}) and tend to be lower in the central parts of the galaxies. The lower central values could be understood by considering Eq.~\eqref{eqn:find_kIR_spatially_resolved}, in which the numerator is only affected by younger stellar populations while the denominator or dust emission is affected by the presence of both young and old stars in the spatial region being probed. Thus, in the central regions of the galaxies, where older populations dominate, there is an increasing contribution of older stars to dust emission, resulting in lower values of $k_{IR}$.

In order to readily compare the $k_{IR}$ values found in this work to that in the literature, we employ the light-weighted mean value, ${\langle k_{IR} \rangle}_{IR}$\footnote{In contrast, the mean $k_{IR}$ is represented simply by ${\langle k_{IR} \rangle}$} for each of $24\,\mu\rm{m}, 70\,\mu\rm{m}, \text{ and } TIR$ bands.
Figure~\ref{fig:k_variation_aperture} considers the impact of inclination and aperture on the ${\langle k_{IR} \rangle}_{IR}$ using all the pixels from the \texttt{Auriga-D} sample. The aperture is reported as a fraction of $R_{\rm{opt}}$, $f_{\text{R,opt}}$, and uses 10 equi-spaced values between $0.25 \times R_{\rm{opt}}$ and $R_{\rm{opt}}$. It is worth noting that the median value of $R_{\rm{opt}}$ for the Auriga sample is $24.25\,\rm{kpc}$. This value is significantly higher than the FOV\footnote{FOV is assumed to be twice the radius of the circular aperture.} of the galaxies in B16. By visual inspection, we find that the FOV of all 8 galaxies used in that work is between $15-20\,\rm{kpc}$. This range with respect to the median Auriga sample's $R_{\rm{opt}}$ is represented by the gray region in Fig.~\ref{fig:k_variation_aperture}.

As anticipated, due to the lower central values, enlarging the aperture size results in a consistent increase in ${\langle k_{IR} \rangle}_{IR}$. Similarly, increasing the inclination also leads to an increase in ${\langle k_{IR} \rangle}_{IR}$, because as the inclination grows, pixels receive more contributions from the outer regions of the galaxies. 

Comparing to the values reported in B16, we consider two apertures $f_{\text{R,opt}}=0.33$, which corresponds to a median Auriga FOV value of $\sim 15$ kpc which is similar to that in B16, and a larger value, $f_{\text{R,opt}}=1$. In the case of the smaller aperture, for an Auriga sample which contains all three inclination used in this work, we find that the ${\langle k_{24} \rangle}_{24}$ as well as the $1\,\sigma$ dispersion are in a very good agreement with similar quantities reported in B16. On the other hand, both ${\langle k_{70} \rangle}_{70}$ and ${\langle k_{TIR} \rangle}_{TIR}$ are nearly $40\,\%$ lower in the case of the smaller aperture. Although, for the same aperture, the ${\langle k_{TIR} \rangle}_{TIR}$ value is lower by about $25\,\%$ with respect to that in L11. 
For the larger aperture, we find that ${\langle k_{24} \rangle}_{24}$, ${\langle k_{70} \rangle}_{70}$, and ${\langle k_{TIR} \rangle}_{TIR}$ values for the Auriga sample are around $20\%$ higher, $25\%$ lower, and $25\%$ lower, respectively, in comparison to the values in B16. The ${\langle k_{TIR} \rangle}_{TIR}$ in this case is in good agreement with the value in H11.
These results are tabulated in Table~\ref{Table:correctionFactor_k_IR_broadband}. 

\subsubsection{$k_{IR}$ and sSFR correlations}
\label{subsubsect:ssFR_vs_kIR}
As mentioned earlier in this paper, the numerator in Eq.~\eqref{eqn:find_kIR_spatially_resolved}
is related to recent star formation, while the denominator is affected by dust heating by both recent and older stellar populations. Assuming a local energy balance, we expect local variations in \(k_{\text{IR}}\) to correlate with the sSFR, provided that the denominator can be linearly decomposed based on heating from each source. As noted by B16 (also see \citet{2007ApJ...657..810D}), 
the linear decomposition of the IR contribution due to heating by old and young populations hinges on the proportionality of the IR emission in a specific band to the intensity of the incident radiation field.  Notably, this applies to the TIR and the 24 $\mu$m band. The explanation is trivial in the case of the TIR. For the 24 $\mu$m and 70 $\mu$m, the correlation between incident radiation intensity and band luminosity could break down if the emission peak due grains in equilibrium passes through the band at intensity levels expected for a given application. In such cases, there is no one-to-one mapping between incident intensity and the band luminosity. 
This intensity threshold is high in the case of the 24 $\mu$m band emission, while it is significantly lower for the 70 $\mu$m band \footnote{typical values for \citet[][Fig.~15 in that work]{2007ApJ...657..810D} model are $\sim10^{5}\,U$ and $\sim50\,U$, respectively, where $U$ is the interstellar radiation field in the solar neighborhood.}. 

Figure~\ref{fig:k_vs_sSFR_aperture=1.0} presents the relation between $k_{IR}$ and the sSFR for the \texttt{Auriga-D} sub-sample using the aperture $f_{\text{R,opt}}=1.0$ at the three face-on inclinations. We reiterate that the sSFR values are calculated using the projected mass maps as discussed in Sect.~\ref{sect:data_products}. 
As expected, the Pearson correlation coefficients (PCC) for the $k_{24}$ and  $k_{TIR}$ exhibit higher values in comparison to $k_{70}$. More interestingly, the values of the PCC using Auriga galaxies for $k_{24}$ (0.57) and $k_{TIR}$ (0.73) are lower than those in B16, which have a value of 0.76 and 0.97, respectively. When using a smaller aperture, $f_{\text{R,opt}}=0.33$ for the Auriga sample, we obtain PCC values of 0.73, 0.63, and 0.74 for $k_{24}$, $k_{70}$, and  $k_{TIR}$, respectively. In the case of $k_{24}$, the rise in dispersion in the $k_{24}$-sSFR relation at higher sSFR values seem to be driven by the rapid change in the value of $k_{24}$ for individual star-forming regions as a function of age as shown in Fig.~\ref{fig:k_variation_individual_TODDLERS}. The lack of very small grains in the dust model employed in \texttt{TODDLERS} (lower size cut-off at $0.03\,\mu \rm{m}$) means that the flux in the $24\,\mu$m band rapid declines as the shells expand (see \citetalias{2023MNRAS.526.3871K}). As for the lower $k_{TIR}$, it could be an indication of the local energy balance being violated such that the IR flux is mostly heated by the stellar population outside the pixel \citet{2018MNRAS.476.1705S}. This hypothesis could be readily tested by applying spatially resolved SED fitting to our Auriga models; however, we plan to explore this in future studies.

\subsection{Resolved MIR-FIR colors of the Auriga sample}
\label{Sec:auriga_MIR-FIR colors}

\begin{figure*}
    \centering
\includegraphics[width=.8\textwidth]{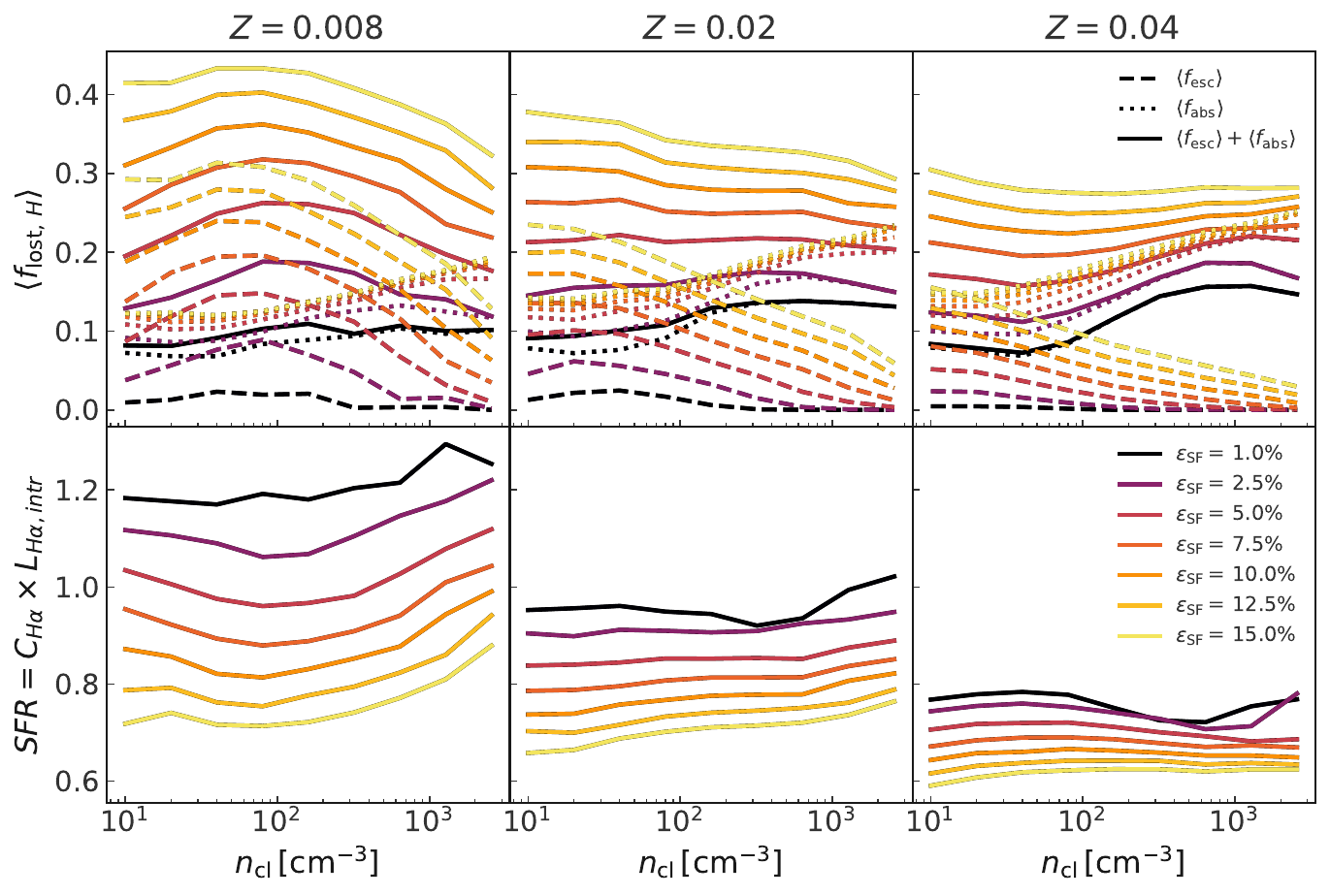}
    \caption{Fraction of LyC photons that do not participate in the production of H$\alpha$ at $Z=0.008,\,0.02, 0.04$ (top) for a population of H\textsc{ii} regions generated (as per Sect.~\ref{sec:recollapse_handling_weights}) for the parameters of the \texttt{TODDLERS} models, $\epsilon_{\rm{SF}}$ and $n_{\rm{cl}}$. The solid line represents the sum of the escape fraction (dashed) and absorption by dust (dotted).  Conversion of H$\alpha$ luminosity to SFR (bottom), using a fixed conversion factor, $C_{H\alpha}=5.5\times 10^{42}$, using the same SF region populations and \texttt{TODDLERS} model parameters as the top row.}
\label{fig:f_esc_f_dust_Z_multi}
\end{figure*}

As discussed in the last section, the shape of the infrared SED is affected by the intensity of incident radiation. This fact is reflected in the color-color relations of the IR SED. Specifically, recent work carried out by \citet[hereafter referred to as G22]{2022ApJ...928..120G} demonstrates that galaxies with higher star-formation rate densities (SFRD) tend to show strong anti-correlations between their MIR (8/24 flux density ratios) and FIR (70/500 flux density ratios) colors. This reinforces the tight color-color relation discussed in \citet[hereafter referred to as C18]{2018ApJ...852..106C}, which was derived from observations of the central starburst region of the dwarf galaxy NGC 4449. Additionally, G22 note that in regions with lower SFRD, the strong MIR-FIR anti-correlation is less evident, primarily due to increased dispersion caused by heating from old stellar populations and the impact of metallicity on  8$\,\mu \rm{m}$ emission. In general, the majority of their galaxies do not exhibit a strong MIR-FIR anti-correlation. 

In the following, we describe how we used the Auriga kpc-scale broadband images and test how they stack against observational results. We have used galaxy morphology as a way to distinguish between galaxies with or without the strong anti-correlation in MIR-FIR colors. Given the large contribution in both MIR and FIR from star-forming regions, we also considered the impact of changing the sub-grid treatment on the color-color relation of Auriga galaxies.

\subsubsection{Relation with morphology}
\label{subsect:auriga_MIRFIR_and_morphology}
Figure~\ref{fig:MIR_FIR_example} shows the color-color relation for three Auriga galaxies, AU-1, AU-2, and AU-21, along with their SFRD maps based on the prescription in L11. These three galaxies encompass the range of MIR-FIR correlations observed in our sample. AU-1 exhibits a relatively high correlation between the blue and red colors of the IR SED, akin to the "Regime-1" galaxies described in G22, which are characterized by higher PCC. Conversely, AU-2 displays considerable dispersion in its data, mirroring the characteristics of NGC5457 or, more broadly, the "Regime-2" galaxies in G22, characterized by the lower correlations between MIR-FIR colors. AU-21 exhibits a behaviour that is between these two. We note that the data points from the Auriga galaxies generally cluster around the relation outlined in C18, whether they exhibit significant dispersion or not.

As noted in G22, the dispersion in the color-color plot largely comes down to the presence of different environments in the same galaxy. Such a difference could be either due to the presence of regions of low SFRD or to a variation in metallicity. This implies that the galaxy morphology could be a predictor of the dispersion in this color-color plane.
In order to tie the dispersion in the color-color plot to the morphology of the galaxies, we used \texttt{statmorph} \citep{2019MNRAS.483.4140R} on the kpc resolution SFRD maps. We used three non-parametric indicators of galaxy morphology (NPIGM) here: concentration, smoothness, and $M_{20}$, which gauge the central concentration, clumpiness, and the presence of non central bright features in a galaxy, respectively. We note that a higher value of smoothness implies higher clumpiness in the image. 

In order to quantify the dispersion on the color-color plot, we use the mean distance of the points from the color-color relation  given in C18, namely, the black dashed curve in Fig.~\ref{fig:MIR_FIR_example}, denoted by $\left< D_{C18} \right>$.
Figure~\ref{fig:Statmorph_SFRmaps} shows $\left< D_{C18} \right>$ as a function of the
three NPIGM for all Auriga galaxies except AU-11, which is undergoing a merger. The points are color-coded by the PCC value obtained for a given galaxy. All three face-on inclinations of the Auriga galaxies have been used to populate Fig.~\ref{fig:Statmorph_SFRmaps}. 

Based on Fig.~\ref{fig:MIR_FIR_example}, galaxies with lower central concentration, higher clumpiness, presence of prominent spiral arms show a higher dispersion, and thus, a lower PCC in the Auriga sample. The smoothness indicator appears to be the best in predicting the dispersion in the MIR-FIR color-color relation at kpc resolutions. This is expected and is in line with the findings in G22 due to the fact that the clumpier systems appear so due to the presence of low and high SFRD regions within the same galaxy. Galaxies with low clumpiness are more uniform in their SFRD maps, and show higher PCC. We note that more compact Auriga galaxies, like AU-10, AU-13, AU-22, AU-26, AU-28, AU-30 exhibit lower dispersion values due their high SFRD.

\subsubsection{MIR-FIR colors of the Auriga sample: \texttt{TODDLERS} vs. \texttt{HiiM3}}

In \citetalias{2023MNRAS.526.3871K}, we examined the MIR-FIR color--color plot employing the a population of \texttt{TODDLERS} star-forming regions. That comparison highlights the difference between \texttt{TODDLERS} and \texttt{HiiM3} libraries in the observational space without considering the effects of the dust exterior to the star-forming regions, which can have a significant effect on the luminosity in the considered bands, particularly in the $8\,\mu \rm{m}$ and $500\,\mu \rm{m}$ bands. Here, we extend that comparison between \texttt{TODDLERS} and \texttt{HiiM3} using Auriga galaxies on a kpc scale. This allows us to visualize the impact of changing the sub-grid treatment from \texttt{TODDLERS} to \texttt{HiiM3} on the color-color plane and compare the results with the observational data from G22.

Figure~\ref{fig:MIRFIR_TODDvsHiiM3} shows the effect of changing the sub-grid treatment of star-forming regions using the \texttt{Auriga-D} sub-sample using the three face-on inclinations. We note that the addition of diffuse dust to the galaxies significantly lowers the $70/500$ color (cf. Fig.~21 in \citetalias{2023MNRAS.526.3871K}). The colors obtained for the Auriga galaxies using \texttt{TODDLERS} are largely consistent with those G22. In contrast, the results for the Auriga galaxies when using \texttt{HiiM3} show an offset with respect to the observational data in both $8/24$ and $70/500$ colors. 
This is because the \texttt{HiiM3} templates exhibit higher $24\,\mu$m and lower $70\,\mu$m luminosities compared to those in the \texttt{TODDLERS} library, using the free parameter values determined by the calibration in K21 (cf. Fig.~\ref{fig:calib_SED}). 

\section{Line-emission based SFR}
\label{sect:emission_line_based_SF_maps}

Next, we used our synthetic line-emission maps to generate SFR maps. We applied two strategies, one using the H$\alpha$ emission and the other using the FIR forbidden lines, \OI$63\,\mu \rm{m}$,  \OIII$88\,\mu \rm{m}$, and \CIIforb$158\,\mu \rm{m}$. We re-gridded all the line-maps used in the upcoming sections to a 1 kpc resolution using the reproject\textunderscore exact algorithm. We note that in order to avoid noisy pixels, we only make use of the pixels at or above the 5$\sigma$ level.

\subsection{SFR from 
\texorpdfstring{H$\alpha$}{} emission}
\label{subsect:Halpha_SFR}

In this section, we concentrate on deriving the SFR from Balmer line maps. 
Having access to both attenuated and transparent maps allows us to calculate a fitting function for the attenuation curve of the H$\alpha$ line emission in Auriga galaxies as a function of the attenuation on resolved scales. We then apply a Balmer decrement-based dust correction to these maps. Additionally, we compare our results with those using 3D radiation-hydrodynamics simulations as reported in \citet[hereafter referred to as T22]{2022MNRAS.513.2904T}

Before delving into the Balmer line maps using the Auriga galaxies at $z=0$, it is worthwhile to look at the predictions from the \texttt{TODDLERS} models, especially at the escape fractions and the LyC destruction due to dust.

\subsubsection{LyC destruction by dust and escape fractions}
\label{subsubsect:LyC_loss}

\begin{figure}
    \centering
    \includegraphics[width=.9\columnwidth]{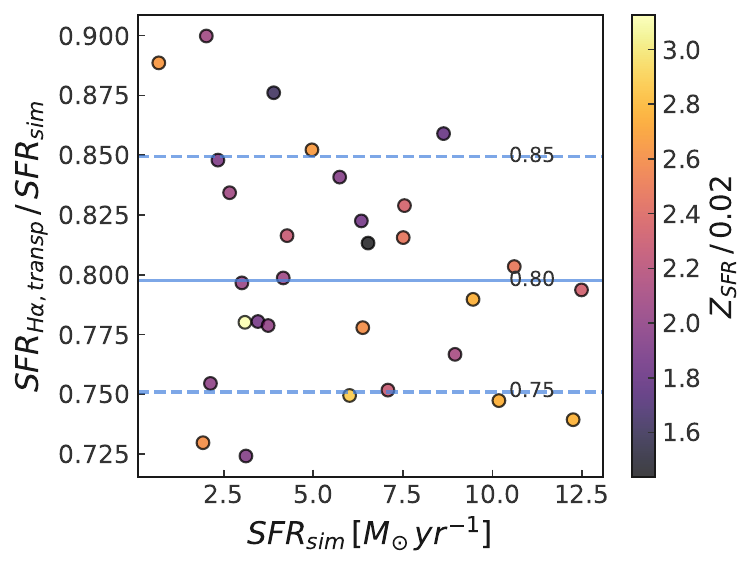}
    \caption{Ratio of the inferred SFR, using a conversion factor of \( C_{H\alpha} = 5.5 \times 10^{42} \), applied to the transparent H\(\alpha\) luminosity, compared to the SFR reported by the simulation across the entire Auriga galaxy sample. The color-coding indicates the SFR-weighted metallicity of each Auriga galaxy.}
    \label{fig:auriga_max_possible_sfr}
\end{figure}

\begin{figure*}
    \centering
\includegraphics[width=.85\textwidth]{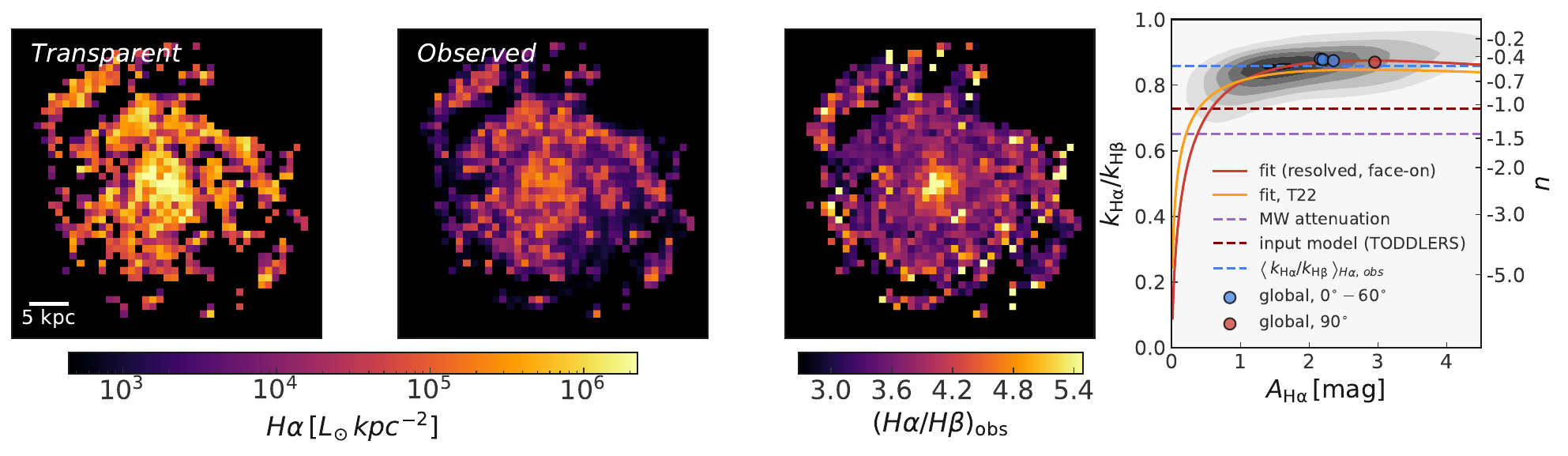}
    \caption{H$\alpha$ maps and attenuation law for one of the Auriga galaxies, AU-1. Panel 1: Transparent H$\alpha$ surface density, Panel 2: Observed H$\alpha$ surface density, Panel 3: The observed Balmer decrement, and Panel 4: The attenuation law found by inverting Eq.~\eqref{eqn:balmer_decrement}, also shown is the fit to the attenuation law for a simulated Milky-Way galaxy done in T22 (orange) and the fit to the resolved face-on data obtained in this work.}
    \label{fig:Halpha_BalmerDec_example}
\end{figure*}

\begin{figure}
    \centering
\includegraphics[width=.9\columnwidth]{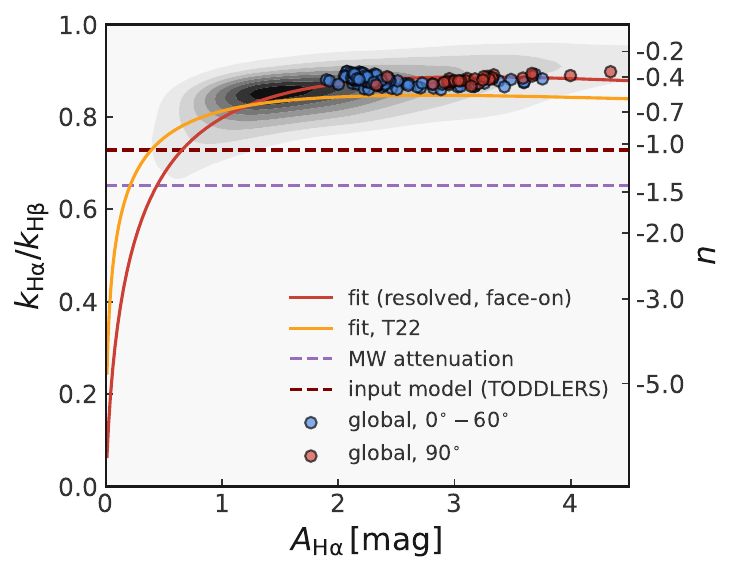}
    \caption{Attenuation law obtained by using the Face-on ($0^{\circ}$ inclination) resolved data for the entire Auriga sample. The circular markers represent the integrated values for Auriga galaxies. The red curve indicates the fit to the data described in Sect.~\ref{subsubsect:dustAttenuation_HiiRegion_and_Correction} while the orange curve is the fit from T22. Also shown are the Milky-Way attenuation (purple dashed line) and the input dust model value employed in \texttt{TODDLERS} models (maroon dashed line).}
    \label{fig:Halpha_atten_all}
\end{figure}

 As discussed in \citetalias{2023MNRAS.526.3871K}, we only employ a mass-based stopping criterion in our \texttt{Cloudy} models, namely, the extent of our models is only dictated by mass, which makes them susceptible to leaking ionizing photons if the shells are rapidly thinned. 
 Apart from this, \texttt{TODDLERS} models consider the presence of dust within \HII regions which leads to the destruction of ionizing photons.
 In both these cases, the amount of ionizing photons that would otherwise lead to the production of H$\alpha$ is reduced.

In order to quantify these effects, we generate a population of star-forming regions with uniform sampled ages between 0-30 Myr and applying the methodology discussed in Sect.~\ref{sec:recollapse_handling_weights} to ensure an SFR of $1\,\rm{M_{\odot}}\,\rm{yr^{-1}}$. 
For this population, we calculate a Hydrogen ionizing photon rate weighted mean escape fraction, using the incident and output spectra of our models in the energy range $1.0-1.8\, \times E_\mathrm{{ion., H}}$, where $E_\mathrm{{ion., H}}$ is the Hydrogen ionizing energy. This averaged quantity, $\left< f_{\mathrm{esc.}}\right>$, is shown with the dashed lines in Fig.~\ref{fig:f_esc_f_dust_Z_multi}.
We similarly define a mean dust absorption fraction, $\left< f_{\mathrm{abs}}\right>$. This is calculated in a manner described in \citetalias{2023MNRAS.526.3871K}. The results for $Z=0.08, \,0.02, \, 0.04$ are shown in the top row of Fig.~\ref{fig:f_esc_f_dust_Z_multi}.

For a population of star-forming regions such as the ones considered here, at the lowest end of the cloud density values, the presence of leaky \HII regions can lead to up to $\approx\,30\,\%$ of the LyC photons escaping the system at $Z=0.02$. This number declines with increasing $n_{\mathrm{cl}}$ and decreasing $\epsilon_{\mathrm{SF}}$. The escape fractions at fixed parameter ($n_{\mathrm{cl}}$, $\epsilon_{\mathrm{SF}}$) generally tend to be lower (higher) at higher (lower) metallicity values shown in Fig.~\ref{fig:f_esc_f_dust_Z_multi}. The escape fractions from the H\textsc{ii} regions are in good agreement with observational studies, such as \citet{2011AAS...21711203C, 2012ApJ...755...40P}, and \cite{2012MNRAS.423.2933R}. We also note that these model escape fractions do not reflect galaxy escape fractions. The photons from leaky H\textsc{ii} regions are expected to be absorbed elsewhere in the ISM and the global escape fraction are not expected to be more than a few percent \citet{2011ApJ...730....5H}. In the current work, we do not model the diffuse gas ionized outside the star-forming regions.

The average fraction of LyC photons absorbed by dust remains between $10-25\%$, and this fraction gradually increases with $n_{\mathrm{cl}}$ and $\epsilon_{\mathrm{SF}}$. This is in good agreement with the values reported in literature \citet{2013seg..book..419C}. On the other hand, based on the radiation-hydrodynamical simulations, T22 report an $\left< f_{\mathrm{abs}} \right>$ of $28\%$ for a Milky-Way like simulation, closer to the higher end of the values encountered in our models. We note that the individual H\textsc{ii} region's $f_{\mathrm{abs}}$ is correlated with the ionization parameter \citepalias{2023MNRAS.526.3871K}, thus, for a population, the averaged value would depend on the SFH. For the population considered here, the variation in the absorbed fraction of LyC photons $\left< f_{\mathrm{abs}} \right>$ across the three metallicities, while keeping other parameters constant, is within $10\%$. 

The bottom row of Fig.~\ref{fig:f_esc_f_dust_Z_multi} shows the estimated SFR using the H$\alpha$ conversion factor for $Z=0.02$ and an IMF suitable for our models ($C_{H\alpha}= 5.5\times10^{-42}$ for H$\alpha$ luminosity in $\rm{erg\,s^{-1}}$).
Figure~\ref{fig:auriga_max_possible_sfr} shows the ratio of the SFR inferred by summing all pixels in the transparent H$\alpha$ map to the simulation value averaged over the past 30 Myr employing the calibrated \texttt{TODDLERS}' parameters. The scatter points' color reflects the SFR weighted metallicity of the Auriga galaxies. It is worth noting that all Auriga galaxies exhibit an SFR weighted mean metallicity lying between $0.03$ and $0.06$. Using the same $C_{H\alpha}$ as above results in inferred to simulation ratio values that lie in the range of 0.725 to 0.9.

\subsubsection{Dust attenuation}
\label{subsubsect:dustAttenuation_HiiRegion_and_Correction}

\begin{figure*}
    \centering
\includegraphics[width=.7\textwidth]{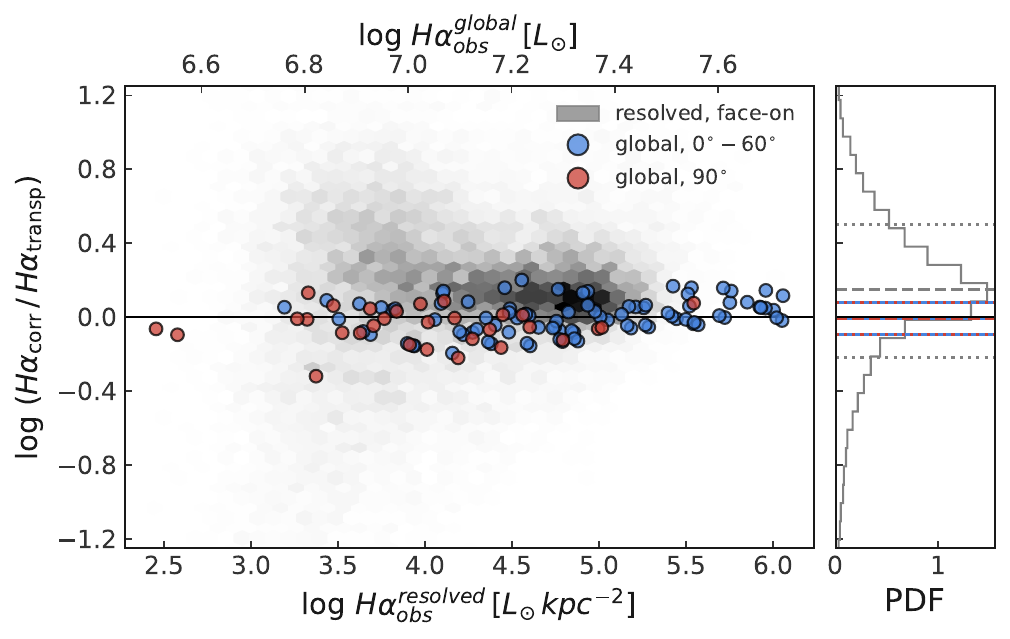}
    \caption{Balmer-decrement based attenuation correction for our models when applied to Auriga galaxies as described in Sect.~\ref{subsubsect:dustAttenuation_HiiRegion_and_Correction}. The ratio of corrected H$\alpha$ to transparent H$\alpha$ is shown as a function of the observed H$\alpha$ luminosity. The circular markers represent the global values while the hexbins represent the kpc-scale data.}
    \label{fig:Balmer_decrement_correction}
\end{figure*}

The wavelength of the H$\alpha$ line makes it highly susceptible to dust attenuation.
Dust attenuation is dealt with using the Balmer decrement correction method, where the theoretical recombination ratio of H$\alpha$ and H$\beta$ line luminosities is known, and any changes to this value are considered to arise from the effects of dust \citep{2006agna.book.....O}. The Balmer decrement could be summarized using the following equation:
\begin{equation}
A_{\rm{H}\alpha} = \frac{k(\lambda_{\rm{H}\alpha})}{k(\lambda_{\rm{H}\beta}) - k(\lambda_{\rm{H}\alpha})} \times 2.5 \log \left( \frac{(\rm{H}\alpha/\rm{H}\beta)_{\text{obs}}}{(\rm{H}\alpha/\rm{H}\beta)_{\text{int}}} \right),
\label{eqn:balmer_decrement}
\end{equation}
where, $A_{\rm{H}\alpha}$ is the H$\alpha$ line attenuation. ${(\rm{H}\alpha/\rm{H}\beta)_{\text{obs}}}$, ${(\rm{H}\alpha/\rm{H}\beta)_{\text{int}}}$ are the observed and intrinsic Balmer decrements, respectively.
$k(\mathrm{H}\alpha)$ and $k(\mathrm{H}\beta)$ represent the extinction curve values for the wavelengths of H$\alpha$ and H$\beta$, respectively. 
The intrinsic Balmer is typically assumed to be 2.87, a Case-B value which holds at an electron temperature of  $10^4\,\rm{K}$ and an electron density of $100\, \rm{cm^{-3}}$ \citet{2006agna.book.....O}. While this ratio can vary if the temperature or density change, this ratio falls in the narrow range of $2.87-3.05$ for a wide variety of conditions, thus for the rest of this work, we used a value of 2.87 for this quantity.

Similar to the approach in T22, given that we have access to both the dust attenuated and transparent H$\alpha$ maps, and thus, $A_{\rm{H}\alpha}$, we invert Eq.~\eqref{eqn:balmer_decrement} to find the attenuation law, or the ratio, $k(\lambda_{\rm{H}\alpha}) / k(\lambda_{\rm{H}\beta})$, as a function of $A_{\rm{H}\alpha}$. An example of this process is given in Fig.~\ref{fig:Halpha_BalmerDec_example}, portraying the intrinsic and attenuated maps required for the calculation of $A_{\rm{H}\alpha}$, and the observed Balmer decrement. Knowing these quantities and the intrinsic Balmer decrement allow us to calculate $k(\lambda_{\rm{H}\alpha}) / k(\lambda_{\rm{H}\beta})$ as a function of $A_{\rm{H}\alpha}$, which is shown in the last column of that figure. 
At this point, it is worth pointing out that our sub-grid models for the Balmer lines, generated using \texttt{Cloudy}, assume isotropic transmission through dusty clouds. Consequently, the lines are subject to dust attenuation as described by the exponential integral function \( E_2(\tau_{d}) \) where $\tau_{d}$ represents the dust optical depth \citep[refer to][for more details on the radiative transfer in \texttt{Cloudy}]{2007A&A...467..187R}. This assumption results in a significantly enhanced intensity decay compared to the exponential decay observed at the same optical depth. 

Following T22, we assume a power-law form of the extinction curve, $k\,\propto\, \lambda^{n}$, and fit the attenuation law using the face-on resolved data by using:
\begin{equation}
\log\left({k_{\rm{H}\alpha}} /{k_{\rm{H}\beta}}\right) = n \log\left({\lambda_{\rm{H}\alpha}}/{\lambda_{\rm{H}\beta}}\right) = a \log\left(A_{\rm{H}\alpha}\right)^2 + b \log\left(A_{\rm{H}\alpha}\right) + c,
\end{equation}
where $a,\, b$, and $c$ are the fitting parameters. The fit to the resolved data from 30 Auriga galaxies is shown in Fig.~\ref{fig:Halpha_atten_all}, with $a=-0.187$, $b=0.183$, and $c=-0.096$.
Additionally, the integrated/global values for the Auriga galaxies are shown as scatter points. The global values follow the fit to the resolved data without any dependence on the inclination. We note that as our modelling does not consider the diffuse H$\alpha$ emission, it misses the lower attenuation data, which results in our fit being lower at $A_{\text{H}\alpha} < 1$ in comparison to the fit in T22.

The median value of $k_{\text{H}\alpha}/k_{\text{H}\beta}$, which defines the slope of the attenuation law, using the integrated $A_{\text{H}\alpha}$ of the Auriga galaxies at all inclinations, as obtained from the resolved attenuation law fit, is $0.88$. This is slightly higher than the value of $0.83$ reported for simulations resembling the Milky-Way in T22. This difference is likely due to the larger median size of the dust grains employed in the \texttt{TODDLERS} model, which is expected to move the attenuation law to shallower slope, \citep[see Fig.~11 in][]{2020ARA&A..58..529S}. Our fit in Fig.~\ref{fig:Halpha_atten_all} lies slightly above the one from T22 for pixels with $A_{\text{H}\alpha} > 2$ for the same reason.

The corrected H$\alpha$ to the transparent H$\alpha$ ratio as a function of the observed H$\alpha$ using the median $k_{\text{H}\alpha}/k_{\text{H}\beta}$ value of $0.88$ is shown in Fig.~\ref{fig:Balmer_decrement_correction}. Using this attenuation law recovers the global H$\alpha$ luminosities within
$\pm\,0.25$ dex. Using the same attenuation law for the resolved data results in a median overestimation of 0.14 dex. This is expected given the dependence of Balmer decrement correction on resolution \citep{2020MNRAS.498.4205V}.

\subsection{SFR from FIR lines}
\label{subsect:FIR_SFR}
\begin{figure*}
    \centering
\includegraphics[width=.85\textwidth]{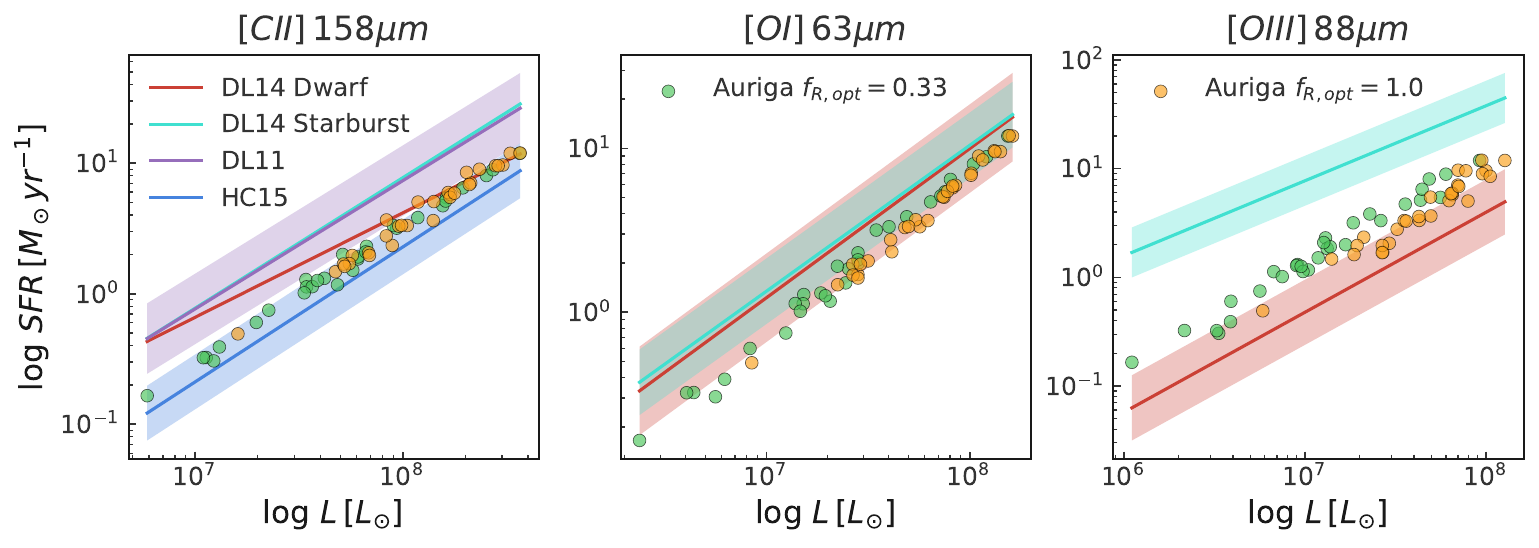}
    \caption{Line luminosity vs. SFR relation for the Auriga galaxies post-processed with the \texttt{TODDLERS} library. The SFR for the Auriga galaxies is the simulation reported value averaged over 30 Myr. The orange points represent values using an aperture (see Sect.~\ref{subsubsect:spatially_varying_kIR}) $f_{\text{R,opt}}=1$, while the green ones are for $f_{\text{R,opt}}=0.33$. The straight lines represent various observational relations and the associated filled areas represent the $1\,\sigma$ dispersion on the calibration.}
    \label{fig:sfr_global_fir_lines}
\end{figure*}

In the relatively cooler regions of the ISM, such as in photo-dissociation regions (PDRs), the emission of far-infrared forbidden lines of elements such as carbon and oxygen facilitates cooling. The first ionization energy of oxygen is $13.62\,\rm{eV}$, which closely matches that of hydrogen. Therefore, in regions of the ISM where Hydrogen is neutral, oxygen is also predominantly O\,\textsc{i}. In contrast, carbon largely exists as the singly ionized \CII due to its lower first ionization energy of $11.3\,\rm{eV}$. This is low enough that the Milky Way's ultraviolet radiation field can maintain the neutral carbon fraction as low as $10^{-3}$ \citep{ryden_pogge_2021}. Given the difference of energies of transition ($E_{\rm{u}}$ in Table~\ref{Tab:fine-structure_lines_data}), \CIIforb158$\,\mu$m (\CIIforb from here onwards) dominates cooling under 100\,K in neutral regions, while \OI63\,$\mu$m  (\OI from here onwards) plays an important role in the $500\leq T\,[\rm{K}]\leq 8000$ neutral regime.
Considering Table~\ref{Tab:fine-structure_lines_data}, \OIII$88\,\mu \rm{m}$ line  (\OIII from here onwards), with an upper state energy of approximately 163\,K and a high ionization potential of 35.1\,eV for O+, emerges primarily from diffuse, highly ionized regions near young O stars. However, due to its low critical density for excitation, other lines, such as \OIII$52\,\mu \rm{m}$ and optical lines including \OIII$5007\,\AA$ and H$\alpha$, might play dominant roles in the cooling of ionized gas regions at higher densities.

The use of \CIIforb158$\,\mu$m as an SFR indicator is directly related to its action as the principal coolant for the neutral atomic gas in the interstellar medium. Consequently, its high luminosity makes it attractive as an indicator for distant galaxies. However, the limitations in terms of density (both electron and hydrogen) in ionized and neutral regimes, respectively, means that \CIIforb line may not effectively trace dense neutral gas that is poised for imminent star formation, or ionized gas around young stars.
\OI serves as an effective coolant in dense and/or warm PDRs. This connection to PDRs links it to star-forming activities. This is also a luminous line, generally observed to be the second brightest line following \CIIforb158$\,\mu$m.
The \OIII line is directly linked to star formation as this line primarily emanates from diffuse, highly ionized regions around young O stars, offering insights into the early stages of star formation, but as mentioned earlier, its low electron critical density could lead to lower brightness.

The \texttt{TODDLERS} library offers support for all of these lines, with the important caveat that these lines only emerge from gas around young stars. It is also worth noting that the results are an aggregate of all star-forming regions, thus, no inclination effects are present.
Future releases of \texttt{SKIRT} will allow users to add contributions of the diffuse media to these lines, it is a worthwhile effort to see the contribution of star-forming regions alone.

 \begin{figure*}
    \centering
\includegraphics[width=.85\textwidth]{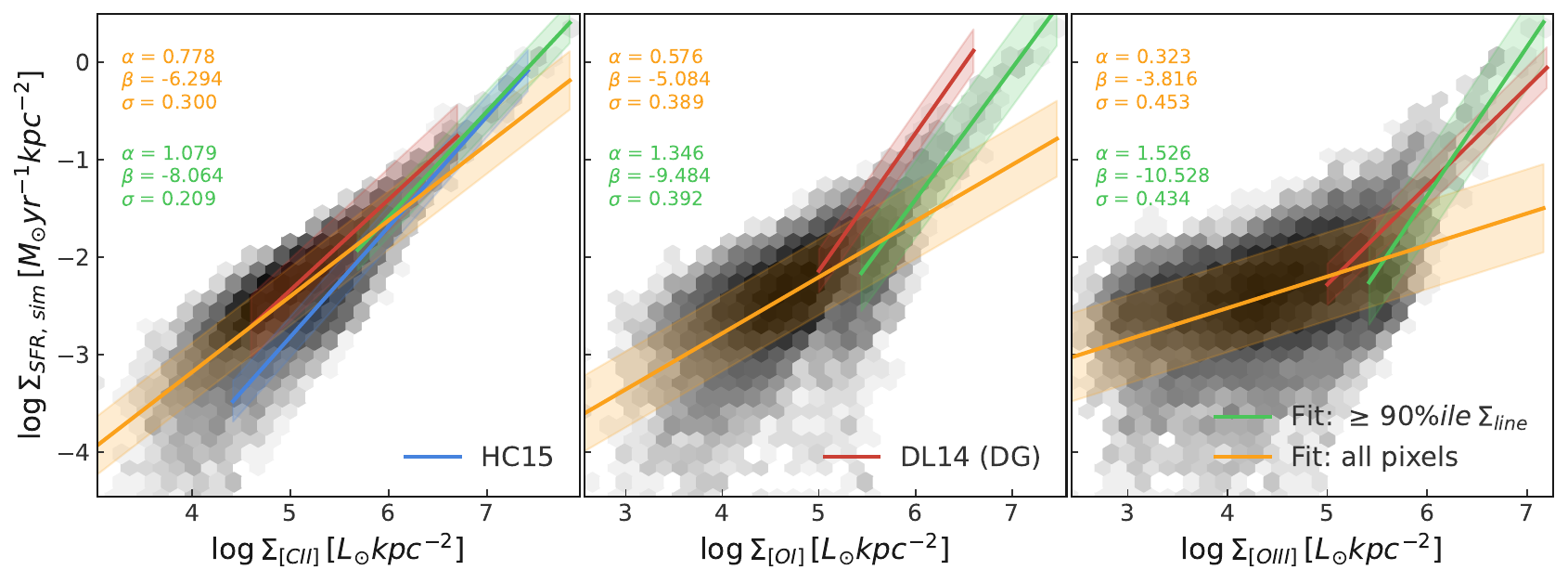}
    \caption{Linear fits in the log-log space  obtained using the line-emission for Auriga galaxies based on Eq.~\eqref{eqn:fine_structure_fit_equation}. The shaded area signifies the $1\,\sigma$ dispersion of the fits. The orange fits uses all the pixels, while the green ones only use the brightest 10\% of pixels out of the entire Auriga sample for each line. The observational fits for metal-poor dwarf galaxies reported in DL14 are shown in red while those for the \CIIforb line reported in HC15 are shown in blue. The filled areas represent the $1\,\sigma$ dispersion on the various fits.}
    \label{fig:sfr_fitAuriga_allLines_comb}
\end{figure*}

\subsubsection{Global SFR comparison}
\label{subsubsect:Global_SFR_comparison_fineStructLines}

\begin{table}[h]
\centering
\begin{tabulary}{\textwidth}{l@{\hskip 0.5cm}c@{\hskip 0.5cm}c@{\hskip 0.5cm}c@{\hskip 0.5cm}c@{\hskip 0.5cm}c}
\toprule
Line & $\alpha$ & $\beta$ & $\sigma~[\text{dex}]$ & Sample & Reference \\
\midrule
\CIIforb & -6.998 & 0.983 & 0.27 &  & DL11 \\ 
\CIIforb & -5.73 & 0.80 & 0.37 & Dwarf & DL14 \\
\CIIforb & -7.06 & 1.0 & 0.27 & Starburst & DL14 \\
\CIIforb & -7.914 & 1.034 & 0.21 &  & HC15 \\
\OI & -6.23 & 0.91 & 0.27 & Dwarf & DL14 \\
\OI & -6.05 & 0.89 & 0.20 & Starburst & DL14 \\
\OIII & -6.71 & 0.92 & 0.30 & Dwarf & DL14 \\
\OIII & -3.89 & 0.69 & 0.23 & Starburst & DL14 \\
\bottomrule
\end{tabulary}
\caption{Fitting function coefficients and dispersion for the global observational data used in this work.}
\label{tab:L-SFR_coefficients_global}
\end{table}

To compare the global SFR values of Auriga galaxies with those in the literature, we reference results from \citet{2011MNRAS.416.2712D}, \citet{2014A&A...568A..62D}, and \citet{2015ApJ...800....1H} (hereafter referred to as DL11, DL14, and HC15, respectively) for the \CIIforb line. For the \OI and \OIII lines, we used results from \citet{2014A&A...568A..62D}. The L11 sample consists of 24 local star-forming galaxies given in \citet{2008ApJS..178..280B}. We used the SFR calibration for the nearby metal poor dwarfs and starburst galaxies reported in DL14 (see Table~3 in that paper). 
The HC15 sample comprises 46 local star-forming galaxies selected from the KINGFISH catalog \citep{2011PASP..123.1347K}. The calibration is given as:
\begin{equation}
    \log (\mathrm{SFR}[\rm{M_{\odot}\, yr^{-1}}])=\alpha \, + \, \beta \,\log (L_{\text {line}}[\rm{L_{\odot}}]),
\end{equation}
with the fitting coefficients for the various observational datasets employed for comparison with our data given in Table~\ref{tab:L-SFR_coefficients_global}.

Figure~\ref{fig:sfr_global_fir_lines} shows the simulations' true SFR averaged over 30 Myr as a function of line luminosity obtained with our post-processing procedure. The $L-$SFR relation is shown for two apertures for the Auriga galaxies. Also shown are the reference literature SFR calibrations from DL11, DL14, and HC15.
The $1\,\sigma$ scatter is shown as filled regions. The $L-$SFR for the Auriga galaxies using the \CIIforb line is in good agreement with the observational data. The effect of aperture does not change the slope of the relation.

As for the $L-$SFR relation for \OI line, while the Auriga galaxies' are largely within the observational scatter, they appear to have somewhat higher \OI line luminosities in comparison to the observed galaxies of similar SFR values. This is likely due to a higher metallicity of our star-forming regions. We expect the self-absorption effects to add additional dispersion in the relation due to inclination effects. The effect of a change in aperture does not lead to a large systematic change in the 
$L-$SFR relation of the Auriga galaxies.

In the case of the \OIII line, the literature calibrations for the dwarf and starburst galaxies show a very large difference. This is expected given the low critical density of the \OIII line. The Auriga  $L-$SFR relation is sandwiched between the relations for the two different galaxy kinds. 
We also note that the starburst sample in DL14 just contains 9 galaxies. The $L-$SFR relation for the smaller aperture shows a systematic offset with respect to the larger one, this is likely an effect of the metallicity gradient in the galaxies. The central parts of the galaxies have a higher metallicity, thus a lower ionizing photon production. Also, the higher mechanical luminosity associated with clusters of higher metallicity leads to shells at higher gas densities.

\subsubsection{Resolved SFR comparison}
\label{subsubsect:resolved_SFR_comparison_fineStructLines}

\begin{table}[h]
\centering
\begin{tabulary}{\textwidth}{l@{\hskip 0.5cm}c@{\hskip 0.5cm}c@{\hskip 0.5cm}c@{\hskip 0.5cm}c}
\toprule
Line & $\alpha$ & $\beta$ & $\sigma~[\text{dex}]$ & Reference \\
\midrule
\CIIforb & -6.99 & 0.93 & 0.32 & DL14 \\
\CIIforb & -8.47 & 1.13 & 0.21 & HC15 \\ 
\OI & -9.19 & 1.41 & 0.22 & DL14 \\ 
\OIII & -7.33 & 1.01 & 0.21 & DL14 \\ 
\bottomrule
\end{tabulary}
\caption{Fitting function coefficients and dispersion for the resolved observational data used in this work.}
\label{tab:L_SFRD_coefficients_resolved}
\end{table}

We employ synthetic maps for the \CIIforb, \OI, and \OIII lines and compare them with the true projected values to create fitting functions for the Auriga galaxies. We fit a straight line in log-log space using the kpc-scale maps of the three fine-structure line maps from all 30 Auriga galaxies. This allows us to determine the values of $\alpha$ and $\beta$ for the equation of the form:
\begin{equation}
\log (\Sigma_{\mathrm{SFR}}[\rm{M_{\odot}\, yr^{-1} \, kpc^{-2}}])=\alpha \, + \, \beta \, \log (\Sigma_{\text {line }}[\rm{L_{\odot}\, kpc^{-2}]})
\label{eqn:fine_structure_fit_equation}
.\end{equation}

The results of this fitting procedure are  are shown in Fig.~\ref{fig:sfr_fitAuriga_allLines_comb}.
Also shown are the  $\Sigma_{\rm{line}}-\Sigma_{\rm{SFR}}$  relations for the low-metallicity dwarf galaxies from DL14. In the case of \CIIforb line, the relation in HC15 is also shown. These relations are given below:

We calculated two kinds of fits for the Auriga galaxies: \begin{enumerate*}
    \item a fit using all pixels with a brightness level at and above 5 $\sigma$ level.
    \item a fit using pixels above the 90 percentile level, namely, the brightest $10\,\%$ pixels.
\end{enumerate*}
Considering the slope of the fits arising from the use of all pixels in the Auriga sample, Auriga galaxies' \CIIforb fits are sub-linear, both DL14 and HC15 report higher values of the slope. Given that we are missing a \CIIforb producing component, it is non-trivial to speculate how its presence would influence this relation. We plan to redo this comparison with a more comprehensive ISM model in future. Interestingly, the $\Sigma_{\CIIforb}-\Sigma_{\rm{SFR}}$ relation for the brightest 10\% of the pixels agrees quite well with the relation given in HC15 as well as DL14.

For $\Sigma_{\OI}-\Sigma_{\rm{SFR}}$ and  $\Sigma_{\OIII}-\Sigma_{\rm{SFR}}$, we find increasing flatter overall relations, only for the most luminous parts of the Auriga galaxies these relations show a super-linear relation. The dispersion in the relation is also higher for the \OI and \OIII lines in comparison to that of the \CIIforb relation with the \OIII line's relation showing the highest dispersion. Remarkably, the slopes of the \OI and \OIII lines' relations of the brightest Auriga pixels show similar slopes as those obtained for the low-metallicity dwarf galaxies in DL14. 

\section{Summary and outlook}
\label{sec:summary}

In this study, we  use the \texttt{TODDLERS} emission library for star-forming regions within \texttt{SKIRT} radiative transfer code to produce several synthetic observations for the 30 Milky Way-like galaxies of the Auriga simulation suite. We have updated the population synthesis method using the \texttt{TODDLERS} library in order to match the SFR of the simulated galaxies while considering the SFR contribution due to recollapsing shells. The calibration of the model parameters follows a method similar to \citet{2021MNRAS.506.5703K}. This involves using analogous galaxies from the DustPedia project and comparing the multi-wavelength broadband data of the simulated galaxies with that of the observed galaxies. The calibrated Auriga broadband data shows higher level of agreement with the DustPedia galaxies, particularly in the UV and MIR bands compared to the data in \citet{2021MNRAS.506.5703K}.

We have produced publicly available broadband and line-emission maps at a spatial resolution of 0.5 kpc. The broadband maps consist of 50 commonly used filters spanning the UV-mm range.
We have calculated two kinds of line-emission maps: \begin{enumerate*}
    \item a set of maps in the optical wavelength range of $0.3-0.7\,\mu \rm{m}$,
    \item a set of maps with specific spectral windows centered at the wavelengths of bright IR lines. These lines are \SIII$18\,\mu \rm{m}$, \SIII$33\,\mu \rm{m}$, \OIII$52\,\mu \rm{m}$, \OI$63\,\mu \rm{m}$, \OIII$88\,\mu \rm{m}$, \NII$122\,\mu \rm{m}$, \CIIforb$158\,\mu \rm{m}$, and \NII$205\,\mu \rm{m}$.
\end{enumerate*} 
Both these have a spectral resolution, $R=3000$. The following summarizes the main results and conclusions of this work:
\begin{itemize}
\item Using the integrated Auriga UV-mm SEDs, we find that the \texttt{TODDLERS} based star-forming regions' treatment allows for higher FUV and NUV attenuation, and lower MIR emission around $24\,\mu \rm{m}$ relative to the SEDs obtained for Auriga galaxies using \texttt{HiiM3}. Earlier studies have consistently pointed out issues with these exact wavelength regimes \citep[e.g.,][]{2019MNRAS.484.4069B, 2020MNRAS.494.2823T, 2021MNRAS.506.5703K, 2022MNRAS.512.2728C, 2022MNRAS.516.3728T, 2024MNRAS.531.3839G}, with SEDs produced with \texttt{HiiM3} exhibiting lower UV attenuation and higher MIR fluxes. Additionally, \texttt{TODDLERS} produces stronger FIR emission in the 50-100$\,\mu \rm{m}$ range, influenced by the non-PAH dust size distribution, which has a lower cut-off at $0.03\,\mu \rm{m}$.

\item We performed a self-consistent calculation of the IR correction factors used for SFR calculation using observed FUV and IR data. This is calculated as:
$k_{IR} = \frac{L_{FUV,\,\rm{intr.}} - L_{FUV}}{L_{IR}}$.
The IR bands utilized for this calculation are 24\,$\mu$m, 70\,$\mu$m, and TIR. 
The $k_{IR}$ values vary significantly across different regions of a galaxy and are influenced by sSFR as dust heating is from both young and old stellar populations. Enlarging the aperture size and increasing galaxy inclination generally increases light-weighted mean $k_{IR}$ values. The $k_{IR}-$sSFR correlations observed in our study were somewhat lower than those reported in \citet{2016A&A...591A...6B}, where the authors employed SED fitting to derive the values. In particular, the lower Pearson correlation coefficient in the case of $k_{TIR}-$sSFR seems to suggest that the local energy balance employed in the SED fitting needs a critical examination. We plan to do this in a future work using our post-processed galaxies and the \texttt{CIGALE} code.

\item Our models show that galaxies with higher SFR densities exhibit strong anti-correlations between their MIR and FIR colors, aligning with observational findings in \cite{2018ApJ...852..106C, 2022ApJ...928..120G}. 
At the same time, the dispersion in MIR-FIR color-color plots is influenced by galaxy morphology, with clumpier systems (higher smoothness values) showing greater dispersion. The presence of low and high SFRD regions within galaxies tends to increase the dispersion through the variations in $8\,\mu \rm{m}$ emission.
It is also noteworthy that the MIR-FIR colors of the Auriga galaxies using the \texttt{TODDLERS} sub-grid treatment were largely consistent with local galaxy data from \citet{2022ApJ...928..120G}. In contrast, the \texttt{HiiM3} templates show significant offsets.

\item Using the \texttt{TODDLERS} library, we quantify the reduction in ionizing photons available for H$\alpha$ production due to \begin{enumerate*}
    \item dust within H\textsc{ii} regions, and 
    \item the escape fractions
\end{enumerate*}. For a population of H\textsc{ii} regions with a constant SFR and $Z \in [0.008, 0.02, 0.04]$, 10-25\% of LyC photons are absorbed by dust, consistent with existing literature. The escape fraction of LyC photons varies with $n_{\rm{cl}}$ and $\epsilon_{\rm{SF}}$, with up to 30\% escaping, decreasing as $n_{\rm{cl}}$ increases and $\epsilon_{\rm{SF}}$ decreases.

\item We have calculated an attenuation law (${k_{\rm{H}\alpha}} /{k_{\rm{H}\beta}}$) for H$\alpha$ based on the dust-attenuated and transparent H$\alpha$ maps. For the kpc-resolved Auriga data, the attenuation law is generally a function of the H$\alpha$ attenuation. We determined a fitting function for the attenuation law as a function of the H$\alpha$ attenuation. The median value of the attenuation law using the global $A_{\rm{H \alpha}}$ and the fitting function is found to be 0.88, which is slightly higher than that reported by the radiation hydrodynamical simulations of \cite{2022MNRAS.513.2904T}. This discrepancy is likely due to the larger median size of dust grains in the \texttt{TODDLERS} models. Applying this attenuation law in the Balmer decrement correction equation corrects the observed global H$\alpha$ luminosities to within $\pm0.25$ dex of the transparent values.

\item Using FIR fine-structure lines \CIIforb 158$\,\mu$m, \OI 63$\,\mu$m, and \OIII 88$\,\mu$m originating from gas in star-forming regions, we compare the results with those in \citet{2011MNRAS.416.2712D, 2014A&A...568A..62D, 2015ApJ...800....1H}. On a global scale, we find a good agreement between the Auriga models and the observational $L-$SFR relation for the \CIIforb line. 
In the case of \OI, the Auriga galaxies exhibit somewhat higher line luminosities at a given SFR compared to results in \citet{2014A&A...568A..62D}, we attribute this to the higher metallicity of the Auriga SF regions. 
The observational \OIII SFR calibration is highly dependent on the galaxy type given the lower critical density of the \OIII line. We find that the Auriga \OIII $L-$SFR relation lies between the starburst and the metal poor dwarf galaxy sample of \citet{2014A&A...568A..62D}. Additionally, we employed synthetic kpc-resolved maps for the \CIIforb, \OI, and \OIII FIR lines and projected SFRD maps to calculate log-log space linear fitting functions for the Auriga galaxies' line brightness and SFRD. Our fits have sub-linear slopes for all three lines when using all the Auriga pixels, but super-linear slopes when using only the brightest 10\% pixels.
\end{itemize}
This work is the first study to use the \texttt{TODDLERS} library for post-processing hydrodynamical simulations. The next step is to use the \texttt{TODDLERS} library to produce high spatial and spectral resolution IFU cubes for a broader array of galaxies (Andreadis et al., in prep.). This effort will utilize \texttt{SKIRT}'s comprehensive support for gas and stellar kinematics. Additionally, we are developing a more detailed ISM model within \texttt{SKIRT} to include emission from diffuse gas. Applying the post-processing methods described in this study to various redshift snapshots will also allow us to directly explore and assess the cosmic evolution of galaxy morphology and various scaling relations. Applying the \texttt{TODDLERS} framework to a broader range of galaxies and across various redshifts will provide valuable insights that can be used to refine the model. In the future, we aim to enhance the model's versatility by adjusting the stellar templates to reflect different abundance sets, the specific physical processes at play, and variations in the IMF. Additionally, we plan to adapt the dust model within \texttt{TODDLERS} to better represent the diverse conditions found in different galactic environments. These modifications will help improve the model's predictions and extend its applicability across a wider range of scenarios.


\section{Data availability}
All the synthetic observations generated for this work are publicly available at \url{www.auriga.ugent.be}.
The cosmological simulation data from the Auriga Project used in this work are publicly available at \url{https://wwwmpa.mpa-garching.mpg.de/auriga/data}.

\begin{acknowledgements}
We thank Remy Indebetouw, Ilse De Looze, Jérémy Chastenet, and Brian Van Den Noortgaete for useful discussions.
We thank Benjamin Gregg for providing the MIR-FIR colors' data from the KINGFISH sample. Additionally, we would like to express our gratitude to the anonymous referee for their valuable feedback. 
MaBa and AG acknowledge the financial support of the Flemish Fund for Scientific Research (FWO-Vlaanderen).
MéBo gratefully acknowledges support from the ANID BASAL project FB210003 and from the FONDECYT regular grant 1211000. This work was supported by the
French government through the France 2030 investment plan managed by the National Research Agency (ANR), as part of the Initiative of Excellence of Université Côte d’Azur under reference number ANR-15-IDEX-01.
The simulations carried out for this work used the Tier-2 facilities of the Flemish Supercomputer Center (https://www.vscentrum.be/) located at Ghent University. 
\end{acknowledgements}


\FloatBarrier
\bibliographystyle{aa}
\bibliography{bibliography}

\FloatBarrier
\appendix

\section{Choice of number photon packets}
\label{appendix:choice of photon packets}

In \citetalias{2021MNRAS.506.5703K}, it was  shown that  $2\times10^{10}$ photon packets offer a good compromise between sufficient signal-to-noise ratio (S/N) and an acceptable simulation run time for image generation in a range of broadbands spanning the UV--submm spectral range. This value was for broadband images generated with Auriga galaxies at 50 pc/pixel spatial resolution. Based on those tests,  we calculate the optimal photon packet number for the various images generated for this work, which have a 500 pc/pixel resolution.
In the case of the broadband images, we have:
\begin{equation}
    N_{\text{ph,BB}} = N_{0} \left( \frac{50}{\Delta x/\text{pc}}\right)^2,
\label{eqn:optimal_photons_broadband}
\end{equation}
where $N_{\text{ph,BB}}$ is the optimal photon packet number for the broadband photometry, $N_{0}=2\times10^{10}$ and
$\Delta x$ is the physical pixel scale in pc.
Equation~\eqref{eqn:optimal_photons_broadband} results in an optimal photon value of $2\times10^{8}$ photons for broadband photometry. 

\begin{figure}
    \centering
\includegraphics[width=\columnwidth]{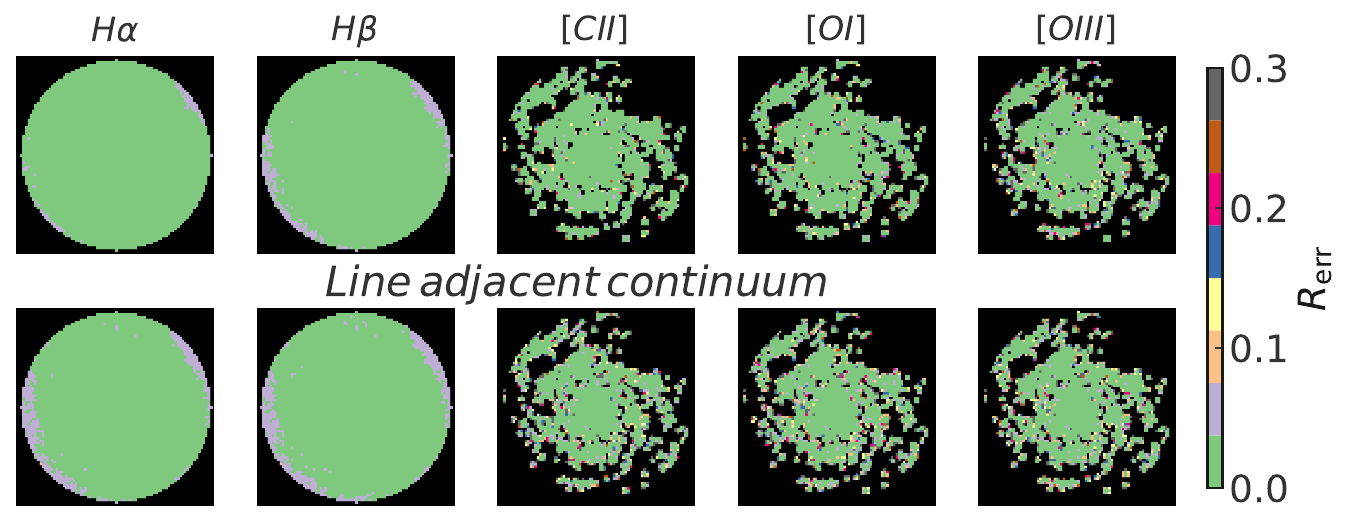}
    \caption{$R_{\rm{err}}$ statistic for the lines and the adjoining continuum at the photon-packet counts given in Sect.~\ref{sect:data_products}. The galaxy shown is AU-1.}
    \label{fig:Rerr_halo1}
\end{figure}

For the high spectral-resolution (HSR) images at $R=3000$, we analyze the photon noise characteristics. We determine that for optical HSR images, a photon count of $N_{\text{ph,HSR}} = 2 \times 10^{11}$ results in a relative error of $R_{\rm{err}} < 0.1$ for the region within $R_{\rm{opt}}$ of the Auriga galaxies. This threshold is required for the results to be considered reliable \citep{2020A&C....3100381C}. In contrast, for the infrared HSR maps, a higher photon count of $N_{\text{ph,HSR}} = 5 \times 10^{11}$ is necessary to achieve a similar level of reliability.
In the case of the infrared HSR maps, where only primary photon-packets are shot, this increase in the number of photon-packets does not impact the run time in a critical way.
Note that these images are always re-gridded to a lower resolution in this work, which increases the signal-to-noise even further.
Figure~\ref{fig:Rerr_halo1} shows the $R_{\rm{err}}$ statistic for one of the Auriga galaxies for the chosen number of photon packets. The relative error remains under 0.05 for most of the pixels.

\section{FIR fine-structure lines' excitation conditions}

Table~\ref{Tab:fine-structure_lines_data} gives the line excitation conditions for the three fine-structure lines used in this work, the second column gives the energy required to ionize the neutral species to the ionization state in question, followed by the energy difference between the levels resulting in the line. The next two columns give the critical Hydrogen density for the lines originating predominantly in neutral regions, and the electron density for those originating in the ionized regions, respectively. Both values are given for \CIIforb158$\,\mu$m, which originates in both regimes. The last column gives the likely sources of origin of the given line. Table adapted from \citet{2014A&A...568A..62D}.
\begin{table}[h]
\centering
\caption{Line excitation conditions for the three fine-structure lines used in this work.}
\begin{tabulary}{\columnwidth}{l@{\hskip 0.2cm}c@{\hskip 0.2cm}c@{\hskip 0.2cm}c@{\hskip 0.2cm}c@{\hskip 0.2cm}>{\raggedright\arraybackslash}p{0.3\columnwidth}}
\toprule
Line & $\mathrm{IE}$ & $E_{\mathrm{u}} / k$ & $n_{\text{crit,H}}$ & $n_{\text{crit,e}}$ & Origin \\
 & $[\mathrm{eV}]$ & $[\mathrm{K}]$ & $[\mathrm{cm}^{-3}]$ & $[\mathrm{cm}^{-3}]$ & \\
\midrule
\CIIforb 158$\mu$m & 11.3 & 91 & $1.6 \times 10^3$ & 44 & PDRs, diffuse H\,\textsc{i} clouds, diffuse ionized gas, \HII regions \\
\OI 63$\mu$m & - & 228 & $5 \times 10^5$ & - & Warm and/or dense PDRs \\
\OIII 88$\mu$m & 35.1 & 163 & - & 510 & Diffuse, highly ionized gas \\
\bottomrule
\end{tabulary}
\label{Tab:fine-structure_lines_data}
\end{table}

\section{Acronyms}
\label{appendix:c}
\begin{figure*}
    \centering
\includegraphics[width=.85\textwidth]{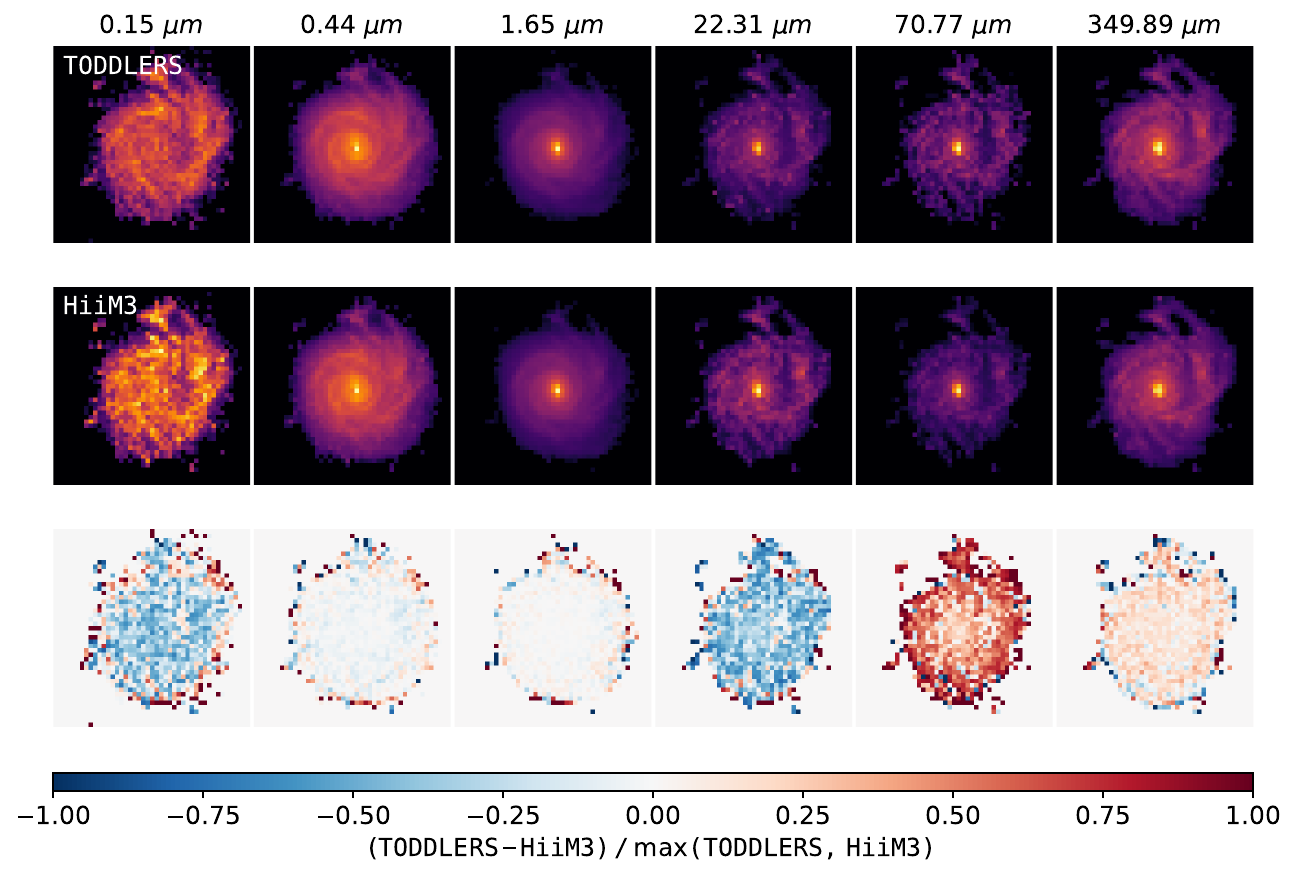}
    \caption{Comparison of spatial maps generated by \texttt{SKIRT} for \texttt{TODDLERS} (top row), \texttt{HiiM3} (middle row) models of AU-21, and the normalized difference between the two (bottom row). Columns show different bands centered at wavelengths listed on top. Images use a cube root scale. For each band, the images for both models share the same colormap limits, but the colormap limits may vary between bands. The data has been smoothed using a uniform filter of three pixels.}
    \label{fig:HiiM3_vs_TODD_images}
\end{figure*}

\begin{table}[h]
\centering
\caption{Acronyms.}
\begin{tabular}{ll}
\hline
Acronym & Definition \\
\hline
BC03 & \citet{2003MNRAS.344.1000B}\\
C18 & \citet{2018ApJ...852..106C} \\
DISM & Dust-containing interstellar medium \\
DL11 & \citet{2011MNRAS.416.2712D} \\
DL14 & \protect{\cite{2014A&A...568A..62D}} \\
\texttt{DPD45} & DustPedia sample of 45 galaxies \\
FIR & Far-infrared \\
FOV & Field of view \\
FUV & Far-ultraviolet \\
G22 & \citet{2022ApJ...928..120G} \\
H11 & \citet{2011ApJ...741..124H} \\
HC15 & \citet{2015ApJ...800....1H} \\
\texttt{HiiM3} & H\textsc{ii} region models from \cite{2008ApJS..176..438G} in \texttt{SKIRT} \\
IFU & Integral field unit \\
IMF & Initial mass function \\
ISM & Interstellar medium \\
K21 & \citet{2021MNRAS.506.5703K} \\
L11 & \citet{2011ApJ...735...63L} \\
LyC & Lyman continuum \\
MIR & Mid-infrared \\
NIR & Near-infrared \\
NPIGM & Non-parametric indicators of galaxy morphology \\
NUV & Near-ultraviolet \\
PAH & Polycyclic aromatic hydrocarbon \\
PCC & Pearson correlation coefficient \\
PDR & Photodissociation region \\
PSF & Point spread function \\
SED & Spectral energy distribution \\
SFH & Star formation history \\
SFR & Star formation rate \\
SFRD & Star formation rate density \\
sSFR & Specific star formation rate \\
SSP & Single stellar population \\
T22 & \citet{2022MNRAS.513.2904T} \\
TIR & Total infrared \\
\hline
\end{tabular}
\label{tab:acronyms}
\end{table}

\end{document}